\documentclass[11pt,a4paper,fleqn]{article}
\usepackage[utf8]{inputenc}
\usepackage{authblk}
\usepackage{geometry}
\usepackage{xcolor}
\usepackage{natbib}
\usepackage{graphicx}
\usepackage{pdflscape}  
\usepackage{appendix}
\usepackage{threeparttable}
\usepackage{booktabs}

\usepackage{amssymb}
\usepackage{amsbsy}
\usepackage{bm}
\usepackage{amsmath}
\usepackage{amsthm}
\usepackage{graphicx}
\usepackage{mathtools}
\usepackage{rotating}
\usepackage{setspace}
\usepackage{enumitem}
\newlist{steps}{enumerate}{1}
\setlist[steps, 1]{label = Step \arabic*:}

\usepackage{dcolumn}
\newcolumntype{d}[1]{D..{#1}} 

\usepackage{caption}
\usepackage{subcaption}
\definecolor{nblue}{HTML}{000660}
\usepackage[colorlinks=true,urlcolor=nblue,linkcolor=nblue,citecolor=nblue]{hyperref}

\title{\textbf{Nowcasting in a Pandemic using Non-Parametric Mixed Frequency VARs}}
\author[a]{Florian \MakeUppercase{Huber}}
\author[b]{Gary \MakeUppercase{Koop}}
\author[c,d]{Luca \MakeUppercase{Onorante}}
\author[a]{\\Michael \MakeUppercase{Pfarrhofer}\thanks{\textit{Corresponding author}: Michael Pfarrhofer. Department of Economics and Salzburg Centre of European Union Studies (SCEUS), University of Salzburg. \textit{Address}: M\"{o}nchsberg 2a, 5020 Salzburg, Austria. \textit{Email}: \href{mailto:michael.pfarrhofer@sbg.ac.at}{michael.pfarrhofer@sbg.ac.at}. Florian Huber and Michael Pfarrhofer gratefully acknowledge financial support from the Austrian Science Fund (FWF, grant no. ZK 35). We would like to thank Serena Ng, Niko Hauzenberger, Paul Hofmarcher, William McCausland, James MacKinnon, Simon van Norden and participants of the Montreal Econometrics Seminar for valuable comments and suggestions. Opinions expressed in this paper do not necessarily reflect the official viewpoint of the European Commission, the European Central Bank, the Oesterreichische Nationalbank or the Eurosystem.}}   
\author[e]{Josef \MakeUppercase{Schreiner}} 
\affil[a]{\textit{University of Salzburg}}
\affil[b]{\textit{University of Strathclyde}}
\affil[c]{\textit{European Commission}}
\affil[d]{\textit{European Central Bank}}
\affil[e]{\textit{Oesterreichische Nationalbank}}
\date{}
\begin{document}

\maketitle\thispagestyle{empty}\normalsize\vspace*{-2em}\small\linespread{1.5}
\begin{center}
\begin{minipage}{0.8\textwidth}
\noindent\small This paper develops Bayesian econometric methods for posterior inference in non-parametric mixed frequency VARs using additive regression trees. We argue that regression tree models are ideally suited for macroeconomic nowcasting in the face of extreme observations, for instance those produced by the COVID-19 pandemic of 2020. This is due to their flexibility and ability to model outliers. In an application involving four major euro area countries, we find substantial improvements in nowcasting performance relative to a linear mixed frequency VAR.
\end{minipage}
\end{center}

\normalsize

\newpage
\section{Introduction}
Mixed frequency vector autoregressions (MF-VARs) have enjoyed great popularity in recent years as a tool for producing timely high frequency nowcasts of low frequency variables. A common practice \citep[see, e.g.,][]{SS}\footnote{A few other recent MF-VAR references adopting similar strategies include \cite{Eraker2015}, \cite{Ghysels2016}, \cite{Braveetal2019} and \cite{KMMP2020}.} is to choose a quarterly macroeconomic variable such as gross domestic product (GDP) and a set of monthly variables and model them together in a VAR so as to produce monthly nowcasts of GDP. The fact that statistical agencies release data such as GDP with a delay, whereas appropriately chosen monthly variables are released with less of a delay further enhances the benefits of the MF-VAR. Nowcasts can be updated in a timely fashion.

The pandemic lockdown of 2020 has further increased the need for timely, high frequency nowcasts of economic activity. And the increasing availability of a variety of high frequency (i.e., monthly, weekly or daily) and quickly released data (i.e., some variables are released almost instantaneously) presents rich opportunities for the mixed frequency modeler. However, the pandemic also poses challenges to the conventional, linear, MF-VAR. During the pandemic, we have seen values of variables that are far from the range of past values. Linear time series econometric methods seek to find average patterns in past data. If current data is very different, using such patterns and linearly extrapolating them may be highly questionable. 

This has led researchers to try to develop new VAR frameworks for nowcasting during the pandemic. For instance, \cite{SSnew} find that the model developed in \cite{SS} nowcasts poorly, but that if they estimate their MF-VAR using data through 2019 and then produce conditional forecasts for the first half of 2020, improvements were obtained. In essence, the extreme data in the first half of 2020 caused estimates of the full sample MF-VAR coefficients to change in a manner which led to poor forecasts. \cite{LenzaPrimiceri} propose an alternative VAR-based approach which allows the error covariance matrix to have a mixture distribution. In essence, the pandemic is treated as a large variance shock and pandemic observations are, thus, drastically downweighted in the model estimation. They conclude: "Our results show that the ad-hoc strategy of dropping these observations may be acceptable for the purpose of parameter estimation. However, disregarding these recent data is inappropriate for forecasting the future evolution of the economy, because it vastly underestimates uncertainty." Thus, although \cite{SSnew} and \cite{LenzaPrimiceri} adopt very different approaches, they end up with similar advice: discard the pandemic observations when estimating the model. 

It is possible to envisage other approaches to modifying the MF-VAR for pandemic times. These would involve parameter change of some form (e.g., structural break or time-varying parameter, TVP, models). But structural break models would be plagued by the fact that there are too few observations post-break to permit reliable estimation. This problem would not occur with TVP models which assume smoothly adjusting coefficents. But such TVP models are not capable of adjusting for sudden and strong jumps in the endogenous variables within a few months such as have been occurring in the pandemic. In light of these considerations, we adopt a different, non-parametric, approach. We argue that such an approach should automatically decide how to treat the pandemic observations in a sensible fashion. In an empirical exercise involving four European countries, we demonstrate the superior nowcasting performance of our approach. 

The non-parametric model we adopt involves Bayesian additive regression trees \citep[BART, see][]{Chipman2010}. BART is a flexible and popular approach in many fields of statistics.\footnote{\cite{BART} is an excellent introduction to BART and includes a long list of papers using BART in a variety of scientific disciplines.} But BART has rarely been used in time series econometrics. \cite{HuberRossini} develop Bayesian methods which build BART into a VAR leading to the Bayesian additive vector autoregressive tree (BAVART) model and demonstrate that it forecasts well. In this paper, we develop Bayesian methods for the mixed frequency version of this model (MF-BAVART). This development is non-trivial and, thus, represents an econometric contribution to the literature even apart from the pandemic context. The MF-VAR is a Gaussian linear state space model and well-established Bayesian methods exist for estimation and predictive inference. However, the MF-BAVART  is not linear and, thus, these methods are not directly available. MF-VARs treat the unobserved high frequency values of the low frequency variables as latent states. Conditional on these latent states, we obtain the BAVART and methods similar to those of \cite{HuberRossini} can be used. It is drawing the latent states (conditional on the BAVART parameters) which is more challenging. We deal with this challenge by rendering the model conditionally Gaussian using recently developed methods for estimating effect sizes in so-called black-box models such as BART, see \citet{Crawfordetal2018,Crawfordetal2019}.\footnote{The methods derived in this literature and used in the present paper can also be used with other black-box models such as neural networks or Gaussian process regressions.} In simulations, we show that this approximation is accurate for data generating processes (DGPs) that  resemble the behavior of GDP during the pandemic.

We apply the resulting model to nowcast GDP growth in selected euro area economies (Germany, Spain, France and Italy) and show that our approach outperforms the linear MF-VAR model. With some exceptions, it produces slightly better nowcasts through 2019. But when the first two quarters of 2020 are included, the improvements offered by MF-BAVART rise substantially. We investigate where these gains come from in a detailed study of the predictive densities for the first six months of 2020. Our results suggest that they stem from the superior ability of our model to adjust the predictive variance in a timely manner and thus increase the probability of observing the extreme observations during the pandemic. These increases in the predictive variance are mostly driven by a flexible Bayesian prior and using a large number of regression trees. Moreover, we also assess whether data on COVID-19 cases and Google mobility trends can improve nowcasts despite the fact that the time series data on these variables is very short. Our model is equipped to extract information from such series and predictive accuracy is increased in some cases.

The remainder of this paper is organized as follows. The next section discusses our econometric methods. We define the MF-BAVART model, illustrate how it can handle extreme observations such as those that have occurred during the 2020 pandemic and discuss prior elicitation. It also contains the relevant full conditional posterior distributions and sketches the  Markov Chain Monte Carlo (MCMC) algorithm for posterior  inference. Section \ref{sec:sim} carries out an artificial data exercise where we investigate the properties of the MF-BAVART model. Section \ref{sec:results} of the paper contains our empirical work. Section \ref{sec:summary} offers a summary and conclusions. Appendix \ref{sec: AppMCMC} provides some technical details. There is also an Online Appendix that includes additional empirical results and MCMC convergence diagnostics.\footnote{All codes and data are available from the corresponding author upon request. Replication files are also available for download at \href{https://github.com/mpfarrho/mf-bavart}{github.com/mpfarrho/mf-bavart}.}

\section{Econometric Methods} \label{sec: econometric_framework}
\subsection{The MF-BAVART}
\label{mfbavart}
Suppose we are interested in modeling an $M$-dimensional vector of time series $\bm y_t = (\bm y'_{m, t}, \bm y'_{q, t})'$  where $\bm y_{m, t}$ is an $M_m$ vector and $\bm y_{q, t}$ is an $M_q$ vector and $t=1, \dots, T$ indicates time at the monthly frequency. The variables in $\bm y_{m, t}$ are observed, but we do not observe $\bm y_{q, t}$ at any point in time. Instead the statistical agency produces a quarterly figure, $\bm y_{Q,t}$. Assuming that $\bm y_{q, t}$ are monthly growth rates (log difference relative to the previous month) and  $\bm y_{Q,t}$ are quarterly growth rates (log difference relative to the previous quarter), the relationship between them is \citep[see][]{Mariano2003}:\footnote{We divide our latent monthly growth rates by three to make their scales comparable. Thus, the right hand side of this equation divides that of \cite{Mariano2003} by three.}
\begin{equation}
\bm y_{Q,t}=\frac{1}{9}\bm y_{q,t}+\frac{2}{9} \bm y_{q,t-1}+\frac{1}{3}\bm y_{q,t-2}+\frac{2}{9} \bm y_{q,t-3}+\frac{1}{9} \bm y_{q,t-4}.
 \label{inter_rest}%
\end{equation}
We refer to this as the intertemporal restriction and note that it applies every third month (e.g., the statistical agency produces quarterly data for the quarter covering January, February and March, but not the quarter covering February, March, April). 

We assume that $\bm y_t$ evolves according to a general multivariate model of the form:
\begin{equation}
    \bm y_t = F(\bm X_t) + \bm \varepsilon_t, \quad \bm \varepsilon_t \sim \mathcal{N}( \bm 0, \bm \Sigma),
\label{nonVAR}    
\end{equation}
with $\bm X_t = (\bm y'_{t-1}, \dots, \bm y'_{t-p})'$ denoting a $K (=Mp)$-dimensional vector of covariates, $F(\bm X_t)=\left(f_1(\bm X_t), \dots, f_M(\bm X_t)\right)'$ being an $M$-dimensional vector of potentially non-linear functions  $f_j: \mathbb{R}^K \to \mathbb{R}$ and $\bm \Sigma$ denotes an $M \times M$-dimensional variance-covariance matrix. 

This is a state space model where unobserved monthly growth rates, $\bm y_{q, t}$, are treated as states. The state equations are given by Eq. (\ref{nonVAR}). The measurement equations are the intertemporal restriction in Eq. (\ref{inter_rest}) (applicable every third month) and those which simply state that $\bm y_{m, t}$ are observed every month. 

If $F(\bm X_t)$ is a vector of linear functions, then we obtain the linear MF-VAR of, e.g., \cite{SS}. Assuming a conditionally Gaussian prior for the VAR coefficients (e.g., the Minnesota prior or a conditionally Gaussian global-local shrinkage prior), posterior and predictive inference is straightforward. That is, standard Bayesian MCMC methods such as Forward-Filtering Backward-Sampling \citep[FFBS, see, e.g.,][]{FS1994} for Gaussian linear state space models can be used. 

In this paper, we wish to treat $F(\bm X_t)$ non-parametrically. In principle, any model can be used for $F$ (e.g., kernel regression, deep neural networks, tree-based models, Gaussian process regression) and the methods derived below could be used with minor modifications. In this paper, we approximate $F$ using BART as, for reasons discussed below, it should be well-designed to capture large shocks and outliers such as those produced by the pandemic. 

BART approximates each $f_j(\bm X_t)$ as follows:
    \begin{equation}
        f_j (\bm X_t) \approx \sum_{s=1}^S g_{js}(\bm X_t| \mathcal{T}_{js}, \bm \mu_{js}),
\label{BART}
    \end{equation}
where $g_{js}$ is a single regression tree function and $\mathcal{T}_{js}$ are the corresponding so-called tree structures related to the $j^{\text{th}}$ element in $\bm y_t$. Moreover,  $\bm \mu_{js}$ are tree-specific terminal nodes and $S$ denotes the total number of trees. The dimension of $\bm \mu_{js}$ is denoted by $b_{js}$ and depends on the complexity of the tree (i.e., this dimension is the number of leaves on the tree). 

The literature on BART models typically sets $S$ between $200$ and $250$. \citet{Chipman2010} show that setting this number too low hurts predictive performance. If $S$ is increased, predictive performance increases up to a certain level of $S$. Increasing the number beyond this level (which is often between $100$ to $150$) hardly impacts predictive accuracy. In our empirical work, including the artificial data exercise, we set $S=250$. This choice provides sufficient flexibility to capture sharp shifts in the conditional mean. Section \ref{subsec:blackbox} contains results showing that this is a reasonable choice.

To understand how BART works, we begin with a single tree and single equation (and, for simplicity, suppress the $j,s$ subscripts which distinguish the various trees and equations in the VAR). In the language of regression, a tree takes as an input the value for the explanatory variables for an observation and produces as an output a fitted value for the dependent variable for that observation. These fitted values are the parameters related to the terminal nodes. It does this by dividing the space of explanatory variables into various disjoint regions using a sequence of binary rules. These so-called splitting rules take the form $\lbrace \bm X \in \mathcal{A}_{r}\rbrace$ or $\lbrace \bm X \not\in \mathcal{A}_{r}\rbrace$ with $\mathcal{A}_{r}$ being a partition set for $r=1,\ldots,b$ and $\bm X = (\bm X_1, \dots, \bm X_T)'$ a full-data matrix of dimension $T \times K$.  The partition rules involve an explanatory variable and depend on whether they are above or below a threshold, $c$. If we let $\bm X_{\bullet i}$ denote the $i^{\text{th}}$ column of $\bm X$, then the partition set takes the form $\lbrace \bm X_{\bullet i} \le c \rbrace$ or $\lbrace \bm X_{\bullet i} > c \rbrace$.
 
For $\bm Y$ (which is a $T$-dimensional vector in our simple example) we obtain the fitted values as follows:
\begin{equation}
    g(\bm X | \mathcal{T}, \bm \mu) = \sum_{r=1}^b \mu_r \mathbb{I}( \bm X \in \mathcal{A}_r), \label{eq: ANOVA}
\end{equation}
which is a step function where $\mathbb{I}$ denotes an indicator function that equals one if its argument is true. As we will show in our simple example in the next sub-section, this is an analysis of variance (ANOVA) model that can, conditional on knowing the indicators, be cast in a linear regression form.

A key point to emphasize is that everything defining the tree is treated as an unknown parameter and estimated. This includes the terminal node parameters ($\bm \mu$ which is the vector of fitted values the algorithm can choose between), their number ($b$) as well as all the elements of the tree structure (i.e. the explanatory variable, $\bm X_{\bullet i}$, and threshold, $c$ chosen to define each splitting rule). To illustrate our model and show what an estimated tree looks like in our dataset, the following sub-section provides an empirical illustration.

\subsection{Empirical Illustration of How BART Works and Handles the Pandemic}
To provide some additional intuition of what BART is doing and why it might be a good approach to handle the extreme observations associated with the pandemic, we preview our empirical application in a simple way. Full details of our data and application are provided below, suffice it to note here that results in this sub-section are for GDP growth in Germany and the full sample of data runs through 2020Q2. We use a single tree with a relatively non-informative prior so as to allow for more complex tree structures. This is just for illustration. In our main empirical work, we use many trees and a regularization prior. 

\begin{figure}[t]
\caption{Estimated tree structures for Germany using a single tree.}
\centering
\begin{subfigure}[t]{\textwidth}
    \caption{Without pandemic observation}\label{tree_normal}
	\includegraphics[width=0.78\textwidth, trim=3cm 0 0 0, clip]{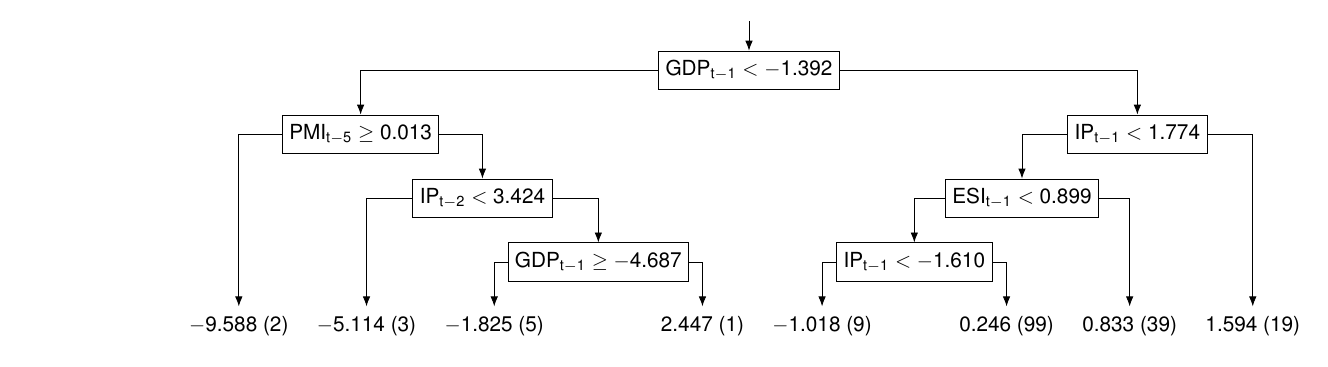}
\end{subfigure}
~
\begin{subfigure}[t]{\textwidth}
    \caption{Including pandemic observation}\label{tree_pandemic}
	\includegraphics[width=0.98\textwidth, trim=4cm 0 0 0, clip]{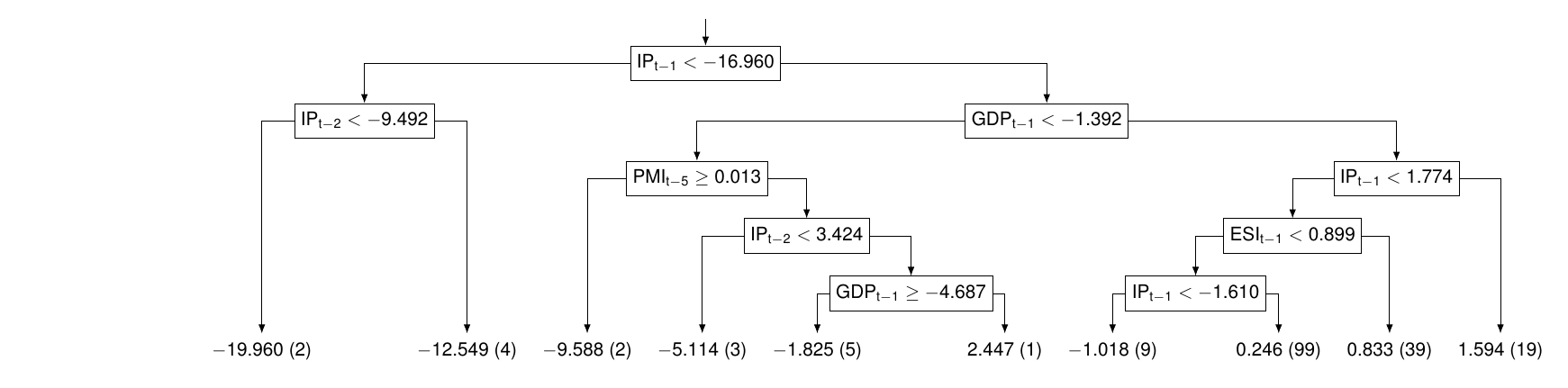}
\end{subfigure}
\begin{minipage}{\linewidth}
\footnotesize \textit{Notes}: The variables in the trees are GDP, IP (industrial production), ESI (economic sentiment indicator), and PMI (purchasing manager's index). Complete definitions are given in the empirical section of this paper. The number of observations choosing each terminal node is in parentheses. The splitting rules are defined such that, if the condition holds you move down the left branch of the tree, else you move down the right branch.
\end{minipage}
\label{fig: DEtree}
\end{figure}

Figure \ref{fig: DEtree} shows two estimated regression trees for Germany. The top panel uses data through $2019$Q$4$ and the bottom panel uses the full sample. The general structure of a regression tree can be illustrated from panel (a). The tree is organized with a condition (e.g., $\text{GDP}_{t-1}<-1.392$) at the top of every binary split. If this condition holds, you move down the left branch, else you move down the right branch. So, for example, the rightmost terminal node ($1.594$) is chosen by observations with $\text{GDP}_{t-1}$ greater than or equal to $-1.392$ (go right at the first split) and have the first lag of industrial production ($\text{IP}_{t-1}$) greater than or equal to $1.774$ (go right at the second split). Hence, the fitted value for GDP growth for observations with last month's industrial production growth greater than $1.774$ and last month's estimated GDP growth above $-1.392$ is $1.594$. There are $19$ observations which fall in this category. 

Recall that everything in the tree is estimated by the algorithm. This includes all the numbers (i.e., the values of the terminal nodes and the thresholds in the splitting conditions), the choice of variables in the splitting conditions (e.g., some of the conditions depend on GDP growth, others depend on the growth in industrial production and both appear at various lags) and the number of splits that occur. For instance, to get to the leftmost terminal node in panel (b) involves checking two conditions (two splits), to get to the rightmost terminal node involves checking three conditions (three splits). To get to some terminal nodes there are multiple splits involving different variables, which is particularly useful for correlated explanatory variables. 

All in all, BART has great flexibility in capturing any sort of behavior, including characteristics common with macroeconomic data. Using Eq. (\ref{eq: ANOVA}), we can illustrate this by casting our model into ANOVA form based on panel (a) of Figure \ref{fig: DEtree}. The corresponding model is given by:
\begin{align*}
    \bm Y_{\bullet j} =  &-9.588~ \lbrace\mathbb{I}(\text{GDP}_{t-1} < -1.392) \mathbb{I}(\text{PMI}_{t-5} \ge 0.013)\rbrace\\ 
    &-5.114~ \lbrace\mathbb{I}(\text{GDP}_{t-1} < -1.392) \mathbb{I}(\text{PMI}_{t-5} < 0.013) \mathbb{I}(\text{IP}_{t-2} < 3.424)\rbrace \\
    &-1.825~\lbrace\mathbb{I}(\text{GDP}_{t-1} < -1.392) \mathbb{I}(\text{PMI}_{t-5} < 0.013) \mathbb{I}(\text{IP}_{t-2} \ge 3.424) \mathbb{I}(\text{GDP}_{t-1} < -4.687)\rbrace\\
    &+ 2.447~ \lbrace\mathbb{I}(\text{GDP}_{t-1} < -1.392) \mathbb{I}(\text{PMI}_{t-5} < 0.013) \mathbb{I}(\text{IP}_{t-2} \ge 3.424) \mathbb{I}(\text{GDP}_{t-1} < -4.687)\rbrace\\
     &- 1.018~\lbrace\mathbb{I}(\text{GDP}_{t-1} \ge -1.392) \mathbb{I}(\text{IP}_{t-1} < 1.774) \mathbb{I}(\text{ESI}_{t-1} < 0.899) \mathbb{I}(\text{IP}_{t-1} < -1.610)\rbrace\\
    &+ 0.246~\lbrace\mathbb{I}(\text{GDP}_{t-1} \ge -1.392) \mathbb{I}(\text{IP}_{t-1} < 1.774) \mathbb{I}(\text{ESI}_{t-1} < 0.899) \mathbb{I}(\text{IP}_{t-1} \ge -1.610)\rbrace\\
    &+ 0.833~\lbrace\mathbb{I}(\text{GDP}_{t-1} \ge -1.392) \mathbb{I}(\text{IP}_{t-1} < 1.774) \mathbb{I}(\text{ESI}_{t-1} \ge 0.899)\rbrace\\
    &+ 1.594~\lbrace\mathbb{I}(\text{GDP}_{t-1} \ge -1.392) \mathbb{I}(\text{IP}_{t-1} \ge 1.774)\rbrace + \bm \varepsilon_{\bullet j}.
\end{align*}
Here we let $\bm Y_{\bullet j}$ and $\bm \varepsilon_{\bullet j}$ the $j^{\text{th}}$ column of $\bm Y$ and $\bm \varepsilon$, respectively. Since GDP is ordered first in $\bm Y$, $j=1$ denotes its equation. Each row in the equation above corresponds to an assignment depending on whether the indicators in the parentheses are equal to one. The equation illustrates that BART is capable of handling multi-way interaction effects. For instance, the first row suggests a two-way interaction effect between the first lag of GDP and the fifth lag of the purchasing managers' index (PMI) while the second row corresponds to a three-way interaction between GDP, PMI and IP.

Another point to note is that our MF-BAVARTs involve six variables, not just the four variables which appear in the estimated trees. The BART algorithm has decided that the other two variables should not be involved in the splitting conditions and have no useful explanatory power for GDP growth (loosely analogous to these other variables being insignificant).

With regards to modeling during the pandemic, we now turn to a comparison of panels (a) and (b) in Figure \ref{fig: DEtree}. Note that panel (b) has a splitting condition $\text{IP}_{t-1}<-16.960$ at the top, an extremely low value for the growth in industrial production. The branch of the tree which satisfies this condition contains six observations. These are the six monthly pandemic observations. The branch of the tree which does not satisfy this condition contains all the non-pandemic observations and is the same as the tree in panel (a). What our tree-based model is doing is creating nodes for capturing outliers. Whereas the parameter estimates in a linear model can be substantially affected by an outlier, BART can simply add a new branch to control for it without affecting the main body of the tree. We will explore this issue in more detail in our empirical work, but this is one of the reasons why our MF-BAVART ends up nowcasting better than the linear MF-VAR, particularly around the time of the pandemic.  

Since this example is a special case with a single tree and thus $S=1$, it ignores another dimension of flexibility in terms of handling outliers: the option to add additional trees. The "A" in BART stands for additive and the BART algorithm allows for new trees to be introduced to deal with the pandemic observations. Through 2019, we could see a tree (or trees) which had terminal nodes covering the range of pre-2020 GDP growth values. For the pandemic period, a new tree could yield an estimate of the average GDP growth rate within the pandemic (which would probably be close to $-10$ percent). Adding such a tree would imply that the remaining tree(s) lack the pandemic-related branches. Thus this new tree could serve to account for deviations from the  pre-pandemic GDP growth rate.

\subsection{The Priors}
In this section we discuss our prior setup. For computational reasons discussed below, all priors are specified in an equation-specific manner and any fixed hyperparameters are the same across equations. In the following discussion, the index $j=1,\dots, M$ refers to the different equations of the model.
 
BART can be interpreted as a non-parametric approach capable of approximating any non-linear function. But, as with any non-parametric approach, BART risks over-fitting. This is why Bayesian methods have been commonly used since prior information can mitigate this problem. We use regularization priors to reduce the complexity of the tree structures and to shrink the terminal nodes. In the jargon of this literature, we force each tree to be small and, thus, act as a weak learner. This essentially implies that for a large $S$, each tree explains only a limited fraction of the variation in $\bm y_t$. For each equation we closely follow \cite{Chipman2010} and use a regularization prior that can be factorized as follows:
\begin{equation*}
p\left( (\mathcal{T}_{j1}, \bm \mu_{j1}), \dots, (\mathcal{T}_{jS}, \bm \mu_{jS}) \right) = \prod_s p(\bm \mu_{js}|\mathcal{T}_{js}) p (\mathcal{T}_{js}).
\end{equation*}
with $p(\bm \mu_{js} | \mathcal{T}_{js}) = \prod_i p(\mu_{i,js} | \mathcal{T}_{js})$ and $\mu_{i, js}$ being the $i^{\text{th}}$ element of $\bm \mu_{js}$.  Within trees, we assume that the terminal leaf parameters are independent of each other but depend on the specific tree structure $\mathcal{T}_{js}$.

We  specify a tree generating stochastic process \citep[see][]{Chipman1998}  on $\mathcal{T}_{js}$ that consists of three parts. The first part relates to the probability that a given node at stage $n = 0, 1, 2, \dots$ is not a terminal node. This probability is specified such that
\begin{equation*}
\alpha (1+n)^{-\beta}.
\end{equation*}
$\alpha \in (0, 1)$ and $\beta > 0$  denote scalar hyperparameters. Smaller (larger) values of $\alpha$  ($\beta$) introduce a larger penalty on more complex tree structures. This prior thus controls for overparameterization by keeping trees rather small and simple (and they thus act as weak learners). In our empirical application we set $\alpha = 0.95$ and $\beta = 2$. This is the standard choice proposed by \cite{Chipman2010} that works well for a wide range of different datasets and in simulations. The second part concerns the possible values the thresholds $c$ can take. Here we assume a discrete uniform distribution over all possible values of the $i^{\text{th}}$ covariate  $\bm X_{\bullet i}$. Finally, the last part deals with the specific variables used in the splitting rules. Again, in the absence of substantial prior information about which variables should define the splitting rules we use a uniform distribution over the $K$ columns of $\bm X$. 
 
It is worth noting that our choice for the tree generating prior flexibly adjusts to the data since the implied prior on the decision rules depends on the range of the columns in $\bm X$. Hence, if $\bm X$ contains extreme observations (such as the ones observed during the pandemic), our prior places equal weights on these extremes without artificially bounding away prior mass from the boundary of the parameter space of the thresholds used to define splitting rules.
 
 On the terminal node parameters we use a Gaussian prior that places substantial prior mass on the range of the $M$ columns in $\bm Y  = (\bm y_1, \dots, \bm y_T)'$. The prior on $\mu_{i, js}$ is given by:
\begin{equation*}
    \mu_{i, js}| \mathcal{T}_{js} \sim \mathcal{N}(0, \sigma^2_{\mu j}).
\end{equation*}
The prior standard deviation $\sigma_{\mu j}$ is set as follows:
\begin{equation}
\sigma_{\mu j} = \frac{\max(\bm Y_{\bullet j})-\min(\bm Y_{\bullet j})}{2 \gamma \sqrt{S}}, \label{eq: prior_MU}
\end{equation}
with $\gamma$ denoting a suitable positive constant. Notice that if the number of trees $S$ or $\gamma$ are increased, the prior is pushed towards zero and the effect of a single tree becomes smaller. This specification also implies that the prior variance widens with the range of $\bm Y_{\bullet j}$. Hence, if we observe extreme observations, the range of $\bm Y_{\bullet j}$ sharply increases and the prior on $\mu_{i, js}$ becomes looser (for fixed $S$ and $\gamma$). This feature helps us in a pandemic since we introduce little shrinkage if outliers arise and consequently allow for a large range of possible realizations of  $\bm Y_{\bullet j}$. And this increases the likelihood of capturing outlying observations when interest centers on predictive inference.
 
Consistent with \cite{Chipman2010}, we construct the prior on $\mu_{i, js}$  by transforming the endogenous variable  such that the transformed values range from $-0.5$ to $0.5$, which implies that the numerator in Eq. (\ref{eq: prior_MU}) is one. Following much of the recent literature, we set $\gamma = 2$. This value implies an approximate $95$ percent probability that the conditional mean of the model is between the minimum and maximum of each column of $\bm Y$.

We specify a prior on the covariance parameters and the error variances $\sigma_j^2$ separately.  Let $\bm Q$ be an $M \times M$ lower triangular matrix with unit diagonal and $\bm H = \text{diag}(\sigma_1^2, \dots, \sigma_M^2)$ denotes a diagonal matrix with elements $\sigma_j^2$, such that $\bm \Sigma = \bm Q \bm H \bm Q'$. The free elements associated with the $j^{\text{th}}$ row of $\bm Q$ are stored in a $(j-1)$-dimensional vector $\bm q_j$.
On each element of  $\bm q_j$ we use a Horseshoe (HS) prior:
\begin{equation}
q_{j i} | \tau_{ji}, \lambda \sim \mathcal{N}(0, \tau_{ji}^2 \lambda^2), \quad \tau_{ji} \sim \mathcal{C}^+(0, 1), \quad \lambda \sim \mathcal{C}^+(0, 1). \label{eq: horseshoe}
\end{equation}
Here we let $C^+$ denote the half Cauchy distribution and $\tau_{ji}$ and $\lambda$ scaling parameters. Note that $\lambda$ does not feature any indices and thus serves as a common shrinkage factor across the free elements of $\bm Q$. For later convenience we let $\underline{\bm V}_j = \lambda^2 \times \text{diag}(\tau^2_{j1}, \dots, \tau^2_{j j-1})$ denote a $(j-1) \times (j-1)$ dimensional prior scaling matrix.

For $\sigma_j^2$ we use the conjugate inverse Chi-square distribution:
\begin{equation*}
\sigma_j^2 \sim \nu_j \xi_j / \chi^2_{\nu_j},
\end{equation*}
where $\nu_j$ and $\xi_j$ denote hyperparameters that are calibrated using a data-based estimate of $\sigma_j$, $\hat{\sigma_j}$. This data-based estimate is taken to be the OLS standard deviation from a univariate AR($5$) model.\footnote{In our real-time nowcasting exercise we compute this quantity on a rolling window basis, always taking into account the available data up to the time that the nowcast is produced.} The values of $\nu_j$ and $\xi_j$ are then chosen such that the $v^{\text{th}}$ quantile of the prior is centered on $\hat{\sigma_j}$ with $P(\sigma_j < \hat{\sigma_j}) = v$. In our application we use $v=0.75$ and set the degrees of freedom $\nu_j=T/2$. We found that this choice avoids too large values of $\sigma^2_j$ during the pandemic and thus forces the tree-based model to fit more aggressively. Smaller values of $\nu_j$ yield similar results if the sample is expanded to include the first two quarters of 2020, but at the cost of potential numerical stability issues of the algorithm.

This completes our prior setup. For reference, Table \ref{tab:priorsummary} summarizes all our prior choices.

\begin{table*}[t]
\caption{Summary of prior hyperparameters.}\vspace*{-1.5em}
\begin{center}
\begin{small}
\begin{threeparttable}
\begin{tabular*}{\textwidth}{@{\extracolsep{\fill}} lll}
 \toprule
 \textbf{Description} & \textbf{Parameters} & \textbf{Hyperparameters}\\
  \midrule
Trees (probability of non-terminal node) & $\alpha(1+n)^{-\beta}$ & $\alpha=0.95$ \\
 & & $\beta=2$\\
Terminal nodes & $\mu_{i,js}|\mathcal{T}_{js}\sim\mathcal{N}(0,\sigma^2_{\mu j})$ & $\sigma_{\mu j}=\left(\max(\bm{Y}_{\bullet j}) - \min(\bm{Y}_{\bullet j})\right)/(2\gamma\sqrt{S})$\\
 & & $\gamma = 2$\\
 & & $S = 250$ (number of trees)\\
Error variances & $\sigma_j^2\sim \nu_j\xi_j/\chi_{\nu_j}^2$ & $\nu_j = T/2$ \\
               & $\xi_j$ s.t. $P(\sigma_j<\hat{\sigma}_j) = v$ & $v=0.75$ \\
Covariances & $q_{ji}|\tau_{ji},\lambda \sim \mathcal{N}(0, \tau_{ji}^2 \lambda^2)$ & $\tau_{ji}\sim\mathcal{C}^{+}(0,1)$ \\
& & $\lambda\sim\mathcal{C}^{+}(0,1)$\\
   \bottomrule
\end{tabular*}
\begin{tablenotes}[para,flushleft]
\scriptsize{\textit{Notes}: ``Description'' provides information about the respective model parameter. ``Parameters'' shows the respective distribution or probability. ``Hyperparameters'' indicates our choice for the hyperparameters. $\bm Y_{\bullet j}$ refers to the $j^{\text{th}}$ column of $\bm Y$, $\hat{\sigma}_j$ is the OLS standard deviation from a univariate AR(5) model and ``s.t.'' is an abbreviation for \textit{such that}. We use this prior setup for the simulation study and our empirical application.}
\end{tablenotes}
\end{threeparttable}
\end{small}
\end{center}
\label{tab:priorsummary}
\end{table*}

\subsection{Posterior Simulation}
In terms of posterior and predictive computation, the point to note is that efficient MCMC algorithms have been derived for estimating BART models. In our MF-BAVART model, we use these conditional on $\bm y_{q, t}$. That is, one block of the MCMC algorithm (to be discussed below) provides draws of $\bm y_{q, t}$ and, conditional on these draws, we use standard algorithms for drawing the BART parameters. In principle, we could draw the parameters of the trees and $\bm \Sigma$ as an entire $M$ dimensional system. However, we follow \cite{Carriero2019} and estimate the model on an equation-by-equation basis by conditioning on the lower Cholesky factor of $\bm \Sigma$. This speeds up computation time enormously.\footnote{Appendix \ref{sec: AppMCMC} shows how to write the model as a system of unrelated regressions (conditional on $\bm \Sigma$).} 
\subsubsection{Drawing the Trees}\label{sec: post_Tree}
In this sub-section, we discuss the sampling step involved in estimating the trees. Each tree structure $\mathcal{T}_{js}$ is obtained using the Bayesian backfitting strategy discussed in \cite{Chipman2010}. Under our prior setup, \cite{Chipman2010} show that the trees can be sampled marginally of $\bm \mu_{js}$:
\begin{equation}
p(\mathcal{T}_{js}|\bm R_{js}, \bm q_j, \bm Z_j, \sigma_j) \propto p(\mathcal{T}_{js}) \underbrace{\int p(\bm R_{js} | \bm \mu_{js}, \mathcal{T}_{js}, \bm q_j, \bm Z_j, \sigma_j) p(\bm \mu_{js}|\mathcal{T}_{js}, \bm q_j, \sigma_j) d \bm \mu_{js}}_{ p(\bm R_{js} |\mathcal{T}_{js}, \bm q_j, \bm Z_j, \sigma_j)}. \label{eq:condpost_tree}
\end{equation}

 We let $\bm R_{js}$ denote a partial residual vector that depends on the trees $\mathfrak{s} \neq s$ as follows:
\begin{equation*}
\bm R_{js} = \bm Y_{\bullet j} - \sum_{\mathfrak{s} \neq s} g_{j \mathfrak{s}}(\bm X | \mathcal{T}_{j \mathfrak{s}}, \bm \mu_{j \mathfrak{s}}) - \bm Z_j \bm q_j,
\end{equation*}
with  $\bm Z_j = (\bm Z_{j1}, \dots, \bm Z_{jT})'$ and $\bm Z_{jt}$ being composed of the structural shocks of the preceding $j-1$ equations (see Appendix \ref{sec: AppMCMC}). Hence, $p(\bm R_{js} |\mathcal{T}_{js}, \bm q_j,\bm Z_j, \sigma_j)$ can be viewed as a conditional likelihood that takes a relatively simple form and, more importantly, does not depend on $\bm \mu_{js}$.

A draw from Eq. (\ref{eq:condpost_tree}) can be obtained by using the Metropolis-Hastings (MH) algorithm proposed in \cite{Chipman1998}. This algorithm starts by generating a candidate value $\mathcal{T}^*_{js}$ from a probability distribution $q(\mathcal{T}^{i}_{js}, \mathcal{T}^*_{js})$, with the superscript $i$ denoting the accepted draw at the $i^{\text{th}}$ iteration of the algorithm. We set $\mathcal{T}^{i+1}_{js} = \mathcal{T}^*_{js}$ with probability:
\begin{equation}
a(\mathcal{T}^{i}_{js}, \mathcal{T}^*_{js}) = \frac{q(\mathcal{T}^*_{js}, \mathcal{T}^{i}_{js}) ~ p(\bm R_{js} |\mathcal{T}^*_{js}, \bm q_j,\bm Z_j , \sigma_j)~ p(\mathcal{T}^*_{js})  }{q(\mathcal{T}^{i}_{js}, \mathcal{T}^*_{js})~ p(\bm R_{js} |\mathcal{T}^i_{js}, \bm q_j,\bm Z_j , \sigma_j)~ p(\mathcal{T}^i_{js})}. \label{eq: acceptance}
\end{equation}

The transition kernel $q(\mathcal{T}^{i}_{js}, \mathcal{T}^*_{js})$ is constructed by switching randomly between four moves.  The first move grows a terminal node with a probability of $0.25$. The second move randomly selects a parent of two terminal nodes and transforms it into a terminal node with a probability of $0.25$.  The third step randomly picks some interior node and changes its splitting rule with probability $0.4$. The final move swaps a decision rule between a parent (i.e., the node above) and child (i.e., the node below) with probability $0.1$. 

The key feature of this algorithm which leads to convenient properties is that $\bm \mu_{js}$ is integrated out and thus the dimension of the estimation problem is kept fixed. 

\subsubsection{Drawing the Latent States and Predictive Inference in the MF-BAVART} \label{sec: sampling_states}
The previous sub-sections defined the MF-BAVART and discussed how well-established MCMC methods can be used to draw the BART parameters conditional on the states (i.e., the unobserved high frequency values of the low frequency variables). To complete the MCMC algorithm we need a method for drawing the states, conditional on the BART parameters. In a linear MF-VAR this is done using standard Bayesian state space algorithms such as FFBS. But with the non-parametric MF-BAVART this is more complicated since the model is highly non-linear and FFBS is not directly applicable. Accordingly, we borrow from the literature that deals with estimating effect sizes in black-box models \citep[see][]{Crawfordetal2018,Crawfordetal2019} to produce a linear approximation to $F(\bm X_t)$. 

This linear approximation has been originally proposed in the context of Gaussian kernel (GK) and Gaussian process (GP) regressions. But the method is much more general and applies to several popular techniques in machine learning and econometrics \citep{Ish_etal}.  The only requirement is that we have learned a non-linear function $\bm F  = (F (\bm X_1), \dots, F(\bm X_T))'$ evaluated at the $T$ observations and we can sample from the relevant posterior distribution.

To explain the linear approximation, note that in linear regression models, the effect size (or regression coefficient) is  interpreted as the magnitude of the projection of $\bm X$ onto $\bm Y$ which takes the form:
    \begin{equation*}
        \hat{\bm A} = \text{Proj}(\bm X, \bm Y) = \bm X^{\dagger} \bm Y,
    \end{equation*}
    with $\bm X^{\dagger}$ (which is $K \times T$) being the Moore-Penrose inverse. In the case where $\bm X$ is a full rank matrix this projection is simply $(\bm X' \bm X)^{-1} \bm X' \bm Y$ and the effect size is the least squares estimate (i.e., it is an estimate of the magnitude of the marginal effect of the explanatory variables on the dependent variables). 
    
    We follow \citet{Crawfordetal2018,Crawfordetal2019} and project $\bm X$ onto the matrix $\bm F$.\footnote{An alternative that might be able to better pick up non-linearities would be to consider non-linear transformations of $\bm X$ such as $\bm X^2$ as well.} This produces the following estimate which can be interpreted as an effect size:
    \begin{equation*}
        \tilde{\bm A} = \text{Proj}(\bm X, \bm F) = \bm X^{\dagger} \bm F.
    \end{equation*}
    The main intuition behind this approximation is that the regression of $\bm F$ on $\bm X$ provides information on how much variance is explained through $\bm X$. This is a simple way of understanding the relationship between $\bm X_{\bullet i}~(i=1, \dots, K)$ on $\bm F$. In the absence of additional regularization, the projection essentially implies that $\bm X \tilde{\bm A} \approx \bm F$.
    
    Given this linear approximation, FFBS can be used to draw $\bm y_{q, t}$. Thus, this step in the MCMC algorithm is an approximate one, but our simulation study and empirical results indicate the approximation is a good one. We use  $\tilde{\bm A}$ to produce a linear approximation to the non-parametric multivariate model:
    \begin{equation}
        \bm y_t = \tilde{\bm A}' \bm X_t + \bm \varepsilon_t.
\label{linearapprox}
    \end{equation}
Since we now have an approximated linear model with Gaussian shocks, standard techniques such as FFBS can be used to draw $\bm y_{q, t}$ based on the Gaussian linear state space model defined by Eq. (\ref{linearapprox}) and the intertemporal restriction in Eq. (\ref{inter_rest}).  

FFBS provides draws from the nowcast distribution based on a linearized version of the non-parametric and non-linear state space model. In our empirical work (and if interest centers on out-of-sample forecasting) we refrain from using these linearized estimates but use the posterior distribution of $\bm F$ to produce now/forecasts. 

In the case of the one-step-ahead forecast, the corresponding predictive distribution is then simply given by:
\begin{equation}
    p(\bm y_{T+1}|\bm X_{T+1})  = \int p(\bm y_{T+1} | \bm \Xi, \bm Y) p(\bm \Xi| \bm Y) d\bm \Xi, \label{eq: pred_dens}
\end{equation}
with $\bm \Xi$ being a generic notation that refers to all parameters and latent states in the model. The conditional distribution $p(\bm y_{T+1} | \bm \Xi, \bm Y)$ is:
\begin{equation}
    \bm y_{T+1} | \bm \Xi, \bm Y \sim \mathcal{N}(F(\bm X_{T+1}), \bm \Sigma). \label{eq: pred_dens_con}
\end{equation}
Within our algorithm, draws from Eq. (\ref{eq: pred_dens_con}) can be mapped back to the quarterly values using the intertemporal restriction in Eq. (\ref{inter_rest}).

Iteratively using this equation allows us to compute higher order forecasts or other functions of the parameters. Equation (\ref{eq: pred_dens}) implies that we can use the BART model to produce the one-step-ahead predictive distribution. After integrating out $\bm \Xi$, this distribution will take a highly non-standard form that allows for multi-modality, fat tails and asymmetries in the forecast distributions. At this point, we would like to stress that for nowcasts, the differences between using the smoothed estimates of $\bm y_{T}$ or using Eq. (\ref{eq: pred_dens_con}) are negligible. Both approaches, after integrating out $\bm \Xi$, yield very similar estimates of the predictive distribution with the same interesting features that make this approach suitable for handling the pandemic.

\subsubsection{Drawing the Remaining Parameters}\label{sec: post_remain}
The steps involved in simulating the remaining parameters are standard with the conditional posteriors taking a well-known form. The terminal node parameters $\bm \mu_{js}$ can be obtained by simulating $\mu_{i,js}$ from independent Gaussian distributions which take a textbook conjugate form. The same can be said about the error variances. These can  be simulated from a conditional posterior which follows an inverse Gamma distribution. 

We sample $\bm q_j$ from a multivariate Gaussian posterior. This posterior is given by:
\begin{equation}
\bm q_j | \star \sim \mathcal{N}\left(\bm m_j, \bm \Omega_j\right),\quad \bm \Omega_j = (\bm Z'_j \bm Z_j + \underline{\bm V}_j^{-1})^{-1}, \quad \bm m_j = \bm \Omega_j \bm Z'_j \tilde{\bm Y}_j \label{eq: q_post}
\end{equation}
with $\tilde{\bm Y}_j  = \bm Y_{\bullet j} - f_j(\bm X)$  and the $\star$ notation indicating that we condition on the remaining model parameters and the latent states.

The scaling parameters of the HS prior are obtained using methods outlined in \cite{makalic2015simple}. Introducing additional auxiliary parameters allows us to simulate $\tau_{ji}$ and $\lambda$  from inverse Gamma distributions. More precisely, the corresponding full conditional posteriors are:
\begin{align*}
\tau^2_{ji}| \star \sim \mathcal{G}^{-1}\left(1, \frac{1}{w_{ji}} + \frac{q_{ji}^2}{2 \lambda^2}\right), \quad \text{ for } i=1, \dots, j-1, \\
\lambda^2 | \star \sim \mathcal{G}^{-1} \left( \frac{M (M-1)+2}{4}, \frac{1}{\zeta} + \frac{1}{2} \sum_i \sum_j\left( \frac{q^2_{ji}}{\tau^2_{ji}}\right)\right).
\end{align*}
The auxiliary parameters $w_{ji}$ and $\zeta$ are simulated from:
\begin{align*}
w_{ji}|\star \sim \mathcal{G}^{-1}(1, 1+\tau_{ji}^{-2}),\quad \zeta | \star \sim \mathcal{G}^{-1}(1, 1+ \lambda^{-2}). 
\end{align*} 

\subsubsection{Full Conditional Posterior Sampling}
Our MCMC algorithm iteratively samples from the full conditional posterior distributions outlined in the previous sub-sections. Conditional on a suitable set of starting values, the algorithm cycles through the following steps:
\begin{enumerate}
\item Simulate the $S$ trees  $\mathcal{T}^{i}_{js}$ for each equation using the Metropolis Hastings algorithm with acceptance probability given by Eq. (\ref{eq: acceptance}) and proposal distribution $q(\mathcal{T}^{i}_{js}, \mathcal{T}^*_{js})$.
\item Simulate the terminal node parameters associated with each tree and for each equation $j$ from a Gaussian distribution. This Gaussian distribution takes a standard form and depends only on the elements in  $\bm Y_{\bullet j} - \bm Z_j \bm q_j$ that are allocated to the respective terminal node and the prior variance.

\item For each equation, simulate the error variance $\sigma_j^2$ from an inverted Gamma distribution.

\item The covariance parameters associated with the $j^{\text{th}}$ row of $\bm Q$ are simulated from a multivariate Gaussian posterior described in Eq. (\ref{eq: q_post}).

\item The parameters related to the HS prior are obtained by sampling from the relevant conditional distributions described above.

\item We use a conventional FFBS algorithm to simulate the latent states based on the approximate model derived in Sub-section \ref{sec: sampling_states}.

\item In case we are interested in producing now/forecasts based on the estimated BART model, use Eq. (\ref{eq: pred_dens_con}) to draw from the predictive distribution.
\end{enumerate}

This algorithm is used to produce $30,000$ draws. We then discard the first $15,000$ draws as burn-in. Standard convergence diagnostics indicate rapid convergence towards the joint posterior distribution and thus closely mirror the excellent performance of the original algorithm of \cite{Chipman2010}. Some of these diagnostics are shown in the supplementary Online Appendix.
   
\section{Simulation Study}\label{sec:sim}
In this section we investigate the properties of our model and algorithm using various DGPs. Since our empirical work deals with nowcasting in the pandemic, we construct DGPs that mirror the dynamic properties observed for actual macroeconomic aggregates such as output and unemployment. We assume that the data is  generated by a persistent VAR(1) process, but during the last part of the sample, we have a substantial deviation from linearity by assuming a rapid decline followed by a sharp and abrupt recovery during the last few observations. To investigate how the quality of our approximation varies with the size of the model, we consider different values for $M \in \{3, 5, 7\}$ as well as different numbers of observations $T \in \{150, 250, 350\}$. 

Assuming that the first element of $\bm y_t$, $y_{1t}$, is the latent process that determines the low frequency variable, our DGP takes the following form:
\begin{equation*}
\bm y_t = \bm A_1 \bm y_{t-1} + \bm Q \bm u_t,\quad \bm \varepsilon_t \sim \mathcal{N}(\bm 0_M, \bm I_M), \text{ for } t=1, \dots, T-25,
\end{equation*}
where $\bm A_1$ is an $M \times M$ matrix with off-diagonal elements sampled from a $\mathcal{N}(0, 0.1^2)$ distribution and its diagonal elements equal to $0.9$. $\bm Q$ is a lower triangular matrix with the diagonal elements equal to $0.2$ and off-diagonal elements sampled from a $\mathcal{N}(0, 0.1^2)$ distribution.\footnote{Since adding random noise to $\bm A_1$ can lead to an unstable DGP with eigenvalues exceeding unity, we introduce the restriction that the maximum absolute eigenvalue of $\bm A_1$ must not exceed one.} The DGP is initialized by simulating $\bm y_0$ from a standard Gaussian distribution.

Non-linearities in the DGP are introduced by assuming that $y_{1t}$ follows a VAR for the first $T-25$ observations. For the remaining $25$ observations we assume that:
\begin{equation*}
y_{1t} = \begin{cases} 
- \vartheta_{0t}  \text{ for } t= T-24, \dots, T-5, \\
+ \vartheta_{1t} \text{ for } t= T-4, \dots, T.
\end{cases}
\end{equation*}
$\vartheta_{jt}~(j=0, 1)$ denotes an element of an evenly spaced grid with the length given by the number of observations within a regime (i.e., $20$ for the downturn and $5$ for the quick recovery). This grid is scaled with the standard deviation of $y_{1t}$, $\sigma_{y_1}$, over the first $T-25$ observations. In the case of the downturn, we let the grid range from $0.1 \sigma_{y_1}$ to $6 \sigma_{y_1}$. In the recovery, the grid runs from $2 \sigma_{y_1}$ to $5 \sigma_{y_1}$. This introduces substantial non-linearities and yields realizations of $y_{1t}$ that behave similar to output during the first half of 2020. It is worth noting that we introduce a larger number of pandemic observations in our simulations than the model gets to see in our empirical work (especially during the downturn). Nevertheless, the rapid increase in $y_{1t}$ takes place over five observations and thus serves to illustrate how our model deals with detecting rapid shifts using few observations.

Notice that the share of these non-linear regimes depends on the length of the time series $T$ and goes from close to seven percent (for $T=350$) to  almost $17$ percent (for $T=150$). These latent indicators are then mapped back to the observed quantities using the intertemporal aggregation scheme outlined in Eq.  (\ref{inter_rest}).

This DGP allows us to assess how well our approach recovers the latent high frequency series $y_{1t}$. Estimation accuracy is measured by computing  root mean squared errors (RMSEs) between the posterior median of $y_{1t}$ and the true outcome.  We compare results from our MF-BAVART specification to a standard linear MF-VAR which is identical in all respects except that it is linear. This implies that we set $F(\bm X_t) = \bm A \bm X_t$ with $\bm A$ being an $M \times K$ coefficient matrix. On $\bm a = \text{vec}(\bm A)$ we use the HS prior defined in Eq. (\ref{eq: horseshoe}). The prior on $\bm \Sigma$ is the same in the two models. All experiments are repeated $150$ times. 

\begin{figure}[t]
\includegraphics[scale=.4]{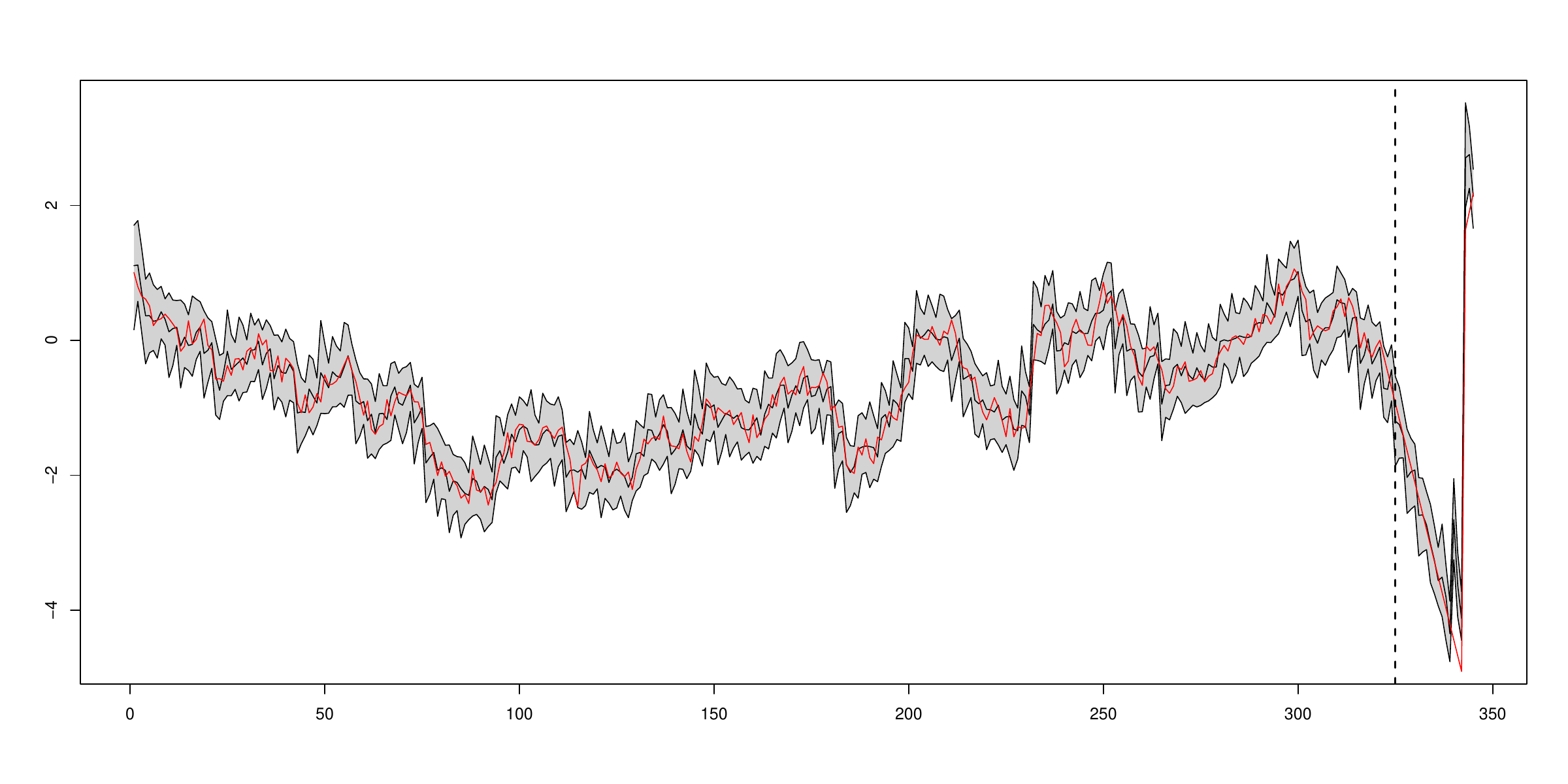}
\caption{Single realization from the DGP for $M=7$ and $T=150$ and posterior estimates}
\caption*{\footnotesize{\textit{Notes}: The gray shaded area and the outer solid  black lines refer to the $5^{\text{th}}$ and $95^{\text{th}}$th credible intervals, the middle solid black line denotes the posterior median and the red line is the true outcome. The dashed black line marks the beginning of the ``crisis'' episode of our DGP.}}\label{fig: DGP1}
\end{figure}

It is worth noting that the chosen DGP is a tough case for our model. This is because the DGP assumes linearity for the majority of periods and then includes a rather small number of extreme observations. Hence, our model needs to learn the behavior of $y_{1t}$ using relatively few data points. And this is the challenge we face when our aim is to model macroeconomic quantities in a pandemic.\footnote{In principle, we could also showcase our model using DGPs that feature substantial non-linearities over the full sample. However, such DGPs imply an unfair advantage of our model relative to the standard MF-VAR. We have carried out some simulation experiments using such DGPs and find that our model performs remarkably well. The results are available from the corresponding author upon request.} 

Before discussing RMSEs comparing our model and the MF-VAR, we present Figure \ref{fig: DGP1} which illustrates the properties of the MF-BAVART model for a single realization from the DGP for $M = 5$ and $T=350$. Other realizations typically look very similar and are thus omitted for brevity. From this figure, we observe that during the first part of the sample (which we label the non-crisis part), our model tracks the actual evolution of $y_{1t}$ remarkably well. Once we hit the crisis regime, it adjusts and the corresponding credible sets include the actual outcome in the vast majority of periods. Only during the beginning of the rapid expansion do the credible sets of MF-BAVART not cover the true series. The final few observations are also well captured with only a single observation not included in the credible intervals. This shows that MF-BAVART works well when the DGP is characterized by substantial non-linearities towards the end of the sample.

To investigate whether this strong performance is consistent across replications of the DGP and how our approach performs relative to the standard MF-VAR, Table \ref{tab: Sims} shows average RMSE ratios between the MF-BAVART and the MF-VAR across different configurations of the DGPs. Values below unity indicate that our model yields more precise estimates of $y_{1t}$ than the competing approach. The figure suggests that for small-scale models ($M=3$), our approach substantially outperforms the linear model for different sample sizes. When the model size is increased, the MF-BAVART still improves upon the MF-VAR but to a slightly lesser extent. This experiment based on synthetic data shows that our approach yields  reasonable estimates of the latent states, even in the presence of substantial deviations from normality and linearity.

\begin{table*}[t]
\caption{Relative performance for differently sized models and sample size.}\vspace*{-1.5em}
\begin{center}
\begin{footnotesize}
\begin{threeparttable}
\begin{tabular*}{\textwidth}{@{\extracolsep{\fill}} lccc}
 \toprule
 &\multicolumn{3}{c}{\textbf{Sample size}}\\
 \cmidrule(lr){2-4}
 \textbf{Model size} & $T=150$ & $T=250$ & $T=350$ \\ 
  \midrule
  $M=3$ & 0.87 & 0.89 & 0.90 \\ 
  $M=5$ & 0.89 & 0.89 & 0.89 \\ 
  $M=7$ & 0.89 & 0.91 & 0.93 \\ 
   \bottomrule
\end{tabular*}
\begin{tablenotes}[para,flushleft]
\scriptsize{\textit{Notes}: The table shows relative root mean squared errors (RMSEs) with respect to the ``true'' latent monthly process between the MF-VAR and MF-BAVART for the estimated monthly processes over $150$ simulated data generating processes (DGPs). Values below unity indicate that our model yields more precise estimates of $y_{1t}$ than the competing approach, and refer to the mean of relative RMSEs over all simulation repetitions.}
\end{tablenotes}
\end{threeparttable}
\end{footnotesize}
\end{center}
\label{tab: Sims}
\end{table*}

\section{Empirical Results}\label{sec:results}
In this section, we investigate the performance of our MF-BAVART model for nowcasting GDP growth using four data sets with relatively short samples. The short sample arises since some of the variables have only been collected for a short time period. This is an issue with many of the new data sets that are becoming popular (e.g., internet search data) and, accordingly, we felt it useful to test our methodology in the type of context where it might be used in the future. All models use a lag length of five since this is the number of lags in the intertemporal restriction in Eq. (\ref{inter_rest}).

\subsection{Data and Design of the Real Time Nowcasting Exercise}
\label{data}
 We use monthly and quarterly data on Germany (DE), France (FR), Italy (IT) and Spain (ES) from 2005M03/2005Q1 to 2020M06/2020Q2 on the following $M = 6$ variables:
 \begin{enumerate}[align=left]
     \item   GDP growth: quarterly GDP growth (abbreviated GDP), released six weeks after the end of the respective quarter. 
            \item Industrial production: monthly growth rate of industrial production (abbreviated IP), released with a lag of approximately six weeks.
            \item Economic sentiment indicator: monthly growth rate of the economic sentiment indicator (abbreviated ESI), released on the next-to-last working day of the respective month.
            \item New car registrations: monthly growth rate of new car registrations (abbreviated CAR), released with a delay of two and a half weeks.
            \item Purchasing managers' index: monthly growth rate of the purchasing managers' index (abbreviated PMI), released on the first working day of the next month.
            \item One-year-ahead interest rates (abbreviated EUR), monthly average of the level of interest rates, available immediately after the end of the respective month.
        \end{enumerate}
Growth rates are exclusively computed by taking log-differences. Data on GDP and industrial production is obtained from the OECD real time database, the Economic Sentiment Indicator is provided by the European Commission, figures on new car registrations are released by the European Automobile Manufacturers Association (ACEA), PMI readings come from Markit and the interest rate data is obtained from Macrobond. In this case the number of columns in $\bm X$ is $K = 30$.

Our main results use the data just described. However, the paper ends with a detailed examination of nowcasting during the COVID-19 pandemic of 2020. One of the models considered there includes two additional variables which we refer to as pandemic variables. The first of these is the number of new COVID-19 infections which is obtained from the Centers for Disease Control. We include this variable as end of period values using the transformation $\log(1+x)$ where $x$ indicates the number of cases. First cases in the respective countries were reported in January 2020. The second variable reflects information associated with social distancing measures and relevant lock-downs by using Google mobility trends data. More precisely, we take the average of the series on Retail \& Recreation, Workplaces and Transit Stations to obtain an aggregate measure of social distancing. Both of these variables are only available for the first six months of 2020. In order to include the pandemic variables in the MF-BAVART we fill in pre-2020 values with zero. For this model, $K$ increases to $K=40$.

Recall that following standard BART practice, our variables are transformed to lie in the interval $[-0.5,0.5]$. This transformation is used for estimating the tree-based components. After obtaining the tree structures and terminal node parameters, the remaining quantities of the model are estimated by transforming the variables back to their original scale.

Given the relatively short sample size we begin evaluating nowcasts in 2011Q1. Within each quarter, we produce three nowcasts, one for each month in the quarter (subsequently abbreviated Mth./Qt.). Our model nowcasts monthly growth rates, $\bm y_{q, t}$, which are turned into quarterly growth rates for comparison with the actual realization of quarterly GDP growth. All of our nowcasts respect the release calendar and data revisions (e.g., a nowcast produced for January will be made at the beginning of February using the data that has been released by then).

\subsection{Results of the Real Time Nowcasting Exercise}
\subsubsection{Comparing the MF-BAVART to the MF-VAR}
\label{compare}
Table \ref{forval2019} summarises our findings for the nowcasting exercise. The table offers a comparison of MF-BAVART to the conventional MF-VAR in terms of RMSEs and log predictive scores (LPSs, which are log predictive likelihoods summed over the nowcast evaluation period). To investigate the pandemic period, we produce two sets of results: one ending in 2019Q4 (the left panel of Table \ref{forval2019}), and one for the full sample (including the pandemic period, the right panel of Table \ref{forval2019}).

\begin{table*}[t]
\caption{Results of the nowcasting exercise.}\vspace*{-1.5em}
\begin{center}
\begin{scriptsize}
\begin{threeparttable}
\begin{tabular*}{\textwidth}{@{\extracolsep{\fill}} ld{1.3}d{2.3}d{1.3}d{3.3}d{1.3}d{4.3}d{1.3}d{4.3}}
  \toprule
   & \multicolumn{4}{c}{\textbf{Through 2019Q4}} & \multicolumn{4}{c}{\textbf{Through 2020Q2}}\\
   \cmidrule(lr){2-5}\cmidrule(lr){6-9}
   & \multicolumn{2}{c}{\textbf{MF-BAVART}} & \multicolumn{2}{c}{\textbf{MF-VAR}} & \multicolumn{2}{c}{\textbf{MF-BAVART}} & \multicolumn{2}{c}{\textbf{MF-VAR}}\\
   \cmidrule(lr){2-3}\cmidrule(lr){4-5}\cmidrule(lr){6-7}\cmidrule(lr){8-9}
& \multicolumn{1}{c}{RMSE} & \multicolumn{1}{c}{LPS} & \multicolumn{1}{c}{RMSE} & \multicolumn{1}{c}{LPS} & \multicolumn{1}{c}{RMSE} & \multicolumn{1}{c}{LPS} & \multicolumn{1}{c}{RMSE} & \multicolumn{1}{c}{LPS}\\
  \midrule
  \textbf{DE} & & & & & & & & \\
Mth./Qt. 1 & 0.586^{\ast\ast} & -54.735^{\ast\ast\ast} & 0.646 & -112.792 & 1.891^{\ast\ast\ast} &  -411.289^{\ast} & 1.971 &  -628.246 \\ 
Mth./Qt. 2 & 0.587 & -54.251^{\ast\ast\ast} & 0.654 & -114.451 & 1.699 &   -87.527^{\ast\ast\ast} & 1.452 &  -182.194 \\ 
Mth./Qt. 3 & 0.537 & -45.714^{\ast\ast\ast} & 0.536 &  -66.582 & 1.610 &   -85.816^{\ast\ast} & 1.001 &  -144.233 \\ 
    \midrule
  \textbf{ES} & & & & & & & & \\
Mth./Qt. 1 & 0.490^{\ast} & -31.535^{\ast\ast\ast} & 0.556 & -151.984 & 3.496 &  -590.363 & 3.421 & -2108.736 \\ 
Mth./Qt. 2 & 0.428^{\ast\ast} & -30.103^{\ast\ast\ast} & 0.484 & -121.163 & 2.828^{\ast\ast} &  -378.179^{\ast\ast} & 2.865 &  -815.515 \\ 
Mth./Qt. 3 & 0.354^{\ast\ast\ast} & -27.170^{\ast\ast\ast} & 0.402 &  -89.600 & 2.814^{\ast\ast\ast} &  -367.837^{\ast} & 2.964 &  -682.207 \\ 
    \midrule
  \textbf{FR} & & & & & & & & \\
Mth./Qt. 1 & 0.320 & -32.317^{\ast\ast\ast} & 0.330 &  -65.015 & 2.669 & -2800.367 & 2.656 & -2893.765 \\ 
Mth./Qt. 2 & 0.312 & -27.547^{\ast\ast} & 0.333 &  -61.902 & 1.991 &  -716.782 & 1.900 & -1122.662 \\ 
Mth./Qt. 3 & 0.313 & -33.201^{\ast} & 0.296 &  -49.014 & 2.057 &  -753.489 & 1.498 & -1141.538 \\ 
    \midrule
  \textbf{IT} & & & & & & & & \\
Mth./Qt. 1 & 0.391^{\ast} & -15.288^{\ast\ast} & 0.473 &  -70.330 & 2.323 &  -360.691^{\ast} & 2.133 &  -872.414 \\ 
Mth./Qt. 2 & 0.366 & -12.075^{\ast\ast} & 0.401 &  -49.272 & 1.808 &  -319.175^{\ast} & 2.153 &  -548.016 \\ 
Mth./Qt. 3 & 0.297^{\ast} &  -6.824^{\ast\ast} & 0.333 &  -26.648 & 1.758 &  -314.247 & 1.558 &  -592.571 \\ 
  \bottomrule
\end{tabular*}
\begin{tablenotes}[para,flushleft]
\scriptsize{\textit{Notes}: The table shows root mean squared errors (RMSE) and cumulative log predictive scores (LPS). Mth./Qt. denotes which month within the quarter the nowcast was made. Asterisks indicate p-values of the \citet{diebold1995comparing} test with MF-VAR as the benchmark model, with levels of significance $^{\ast}$ ($10\%$), $^{\ast\ast}$ ($5$\%), $^{\ast\ast\ast}$ ($1$\%).}
\end{tablenotes}
\end{threeparttable}
\end{scriptsize}
\end{center}
\label{forval2019}
\end{table*}

Note first that, as we move from month to month within a quarter, our nowcasts almost always improve. This statement holds true for both nowcast evaluation metrics and countries. This provides evidence that mixed frequency methods are useful for nowcasting in these data sets. As new information is released each month, our nowcasts of GDP growth improve. 

In terms of the comparison of linear versus non-parametric mixed frequency methods, prior to the pandemic, we find that MF-BAVART nowcasts better than the linear MF-VAR. 
If we consider the properties of our density nowcasts using LPS as the evaluation metric, the superior performance of MF-BAVART holds for each month within the quarter and for every country. The nowcast improvements provided by MF-BAVART are statistically significant using the \citet[DM,][]{diebold1995comparing} test.\footnote{We would like to stress that the DM test provides only a rough measure of statistical accuracy. This is because we use an expanding estimation sample and the number of observations in the hold-out period is quite low. Moreover, the DM test is conservative when applied to short-term forecasting problems.}  For point nowcast performance, RMSEs indicate that MF-BAVART is outperforming the MF-VAR, but the nowcast improvements tend not to be statistically significant. 

When we turn to the right panel of Table \ref{forval2019}, which includes the pandemic period, we tend to see even larger improvements in the nowcast performance of the MF-BAVART relative to the MF-VAR. Note in particular the huge improvements found using LPS (a measure based on the entire predictive density) which occur for every country and for every month within the quarter. In the right panel of the table, differences in LPS between the MF-BAVART and the MF-VAR are typically measured in the hundreds whereas in the left panel they are typically measured in the tens. Given the relationship between log predictive likelihoods and marginal likelihoods, these findings constitute extremely strong evidence in favor of the MF-BAVART.  Thus, standard Bayesian model comparison methods are showing strong evidence that MF-BAVART is handling the pandemic better than the MF-VAR. Assessing statistical difference by means of DM tests points towards superior predictive capabilities in terms of LPS at the $10$ percent significance level for the majority of countries (except France) and months within the quarter.

\subsubsection{A Deeper Examination of why MF-BAVART is Forecasting so Well}\label{subsec:blackbox}
RMSEs are not as consistently favorable to the MF-BAVART as LPSs are when the pandemic observations are included. This indicates that the benefits of using the non-parametric approach lie largely in its ability to better model second and higher predictive moments for the extreme observations for the first half of 2020. More precisely, once we move into the pandemic period our model yields wider predictive intervals. And it produces these wider predictive distributions in a timely manner.

To explain why BART is producing these features, note that BART handles outliers either through creating a new terminal node or by adding trees. The corresponding nodes (either in an existing complicated tree or a newly grown tree) will feature very few observations (often only one or two). This implies little information in the likelihood which is coupled with a large prior variance (mainly driven by our data-based prior discussed in Section \ref{sec: econometric_framework}) thus leading to a large  posterior variance.

\begin{figure}[!ht]
\caption{Posterior interval width of latent GDP versus leverage.}\label{fig: leverage}
\centering
\begin{subfigure}[t]{.45\textwidth}
    \caption{DE}\label{pred_2020_GE_lev}
	\includegraphics[scale=0.35]{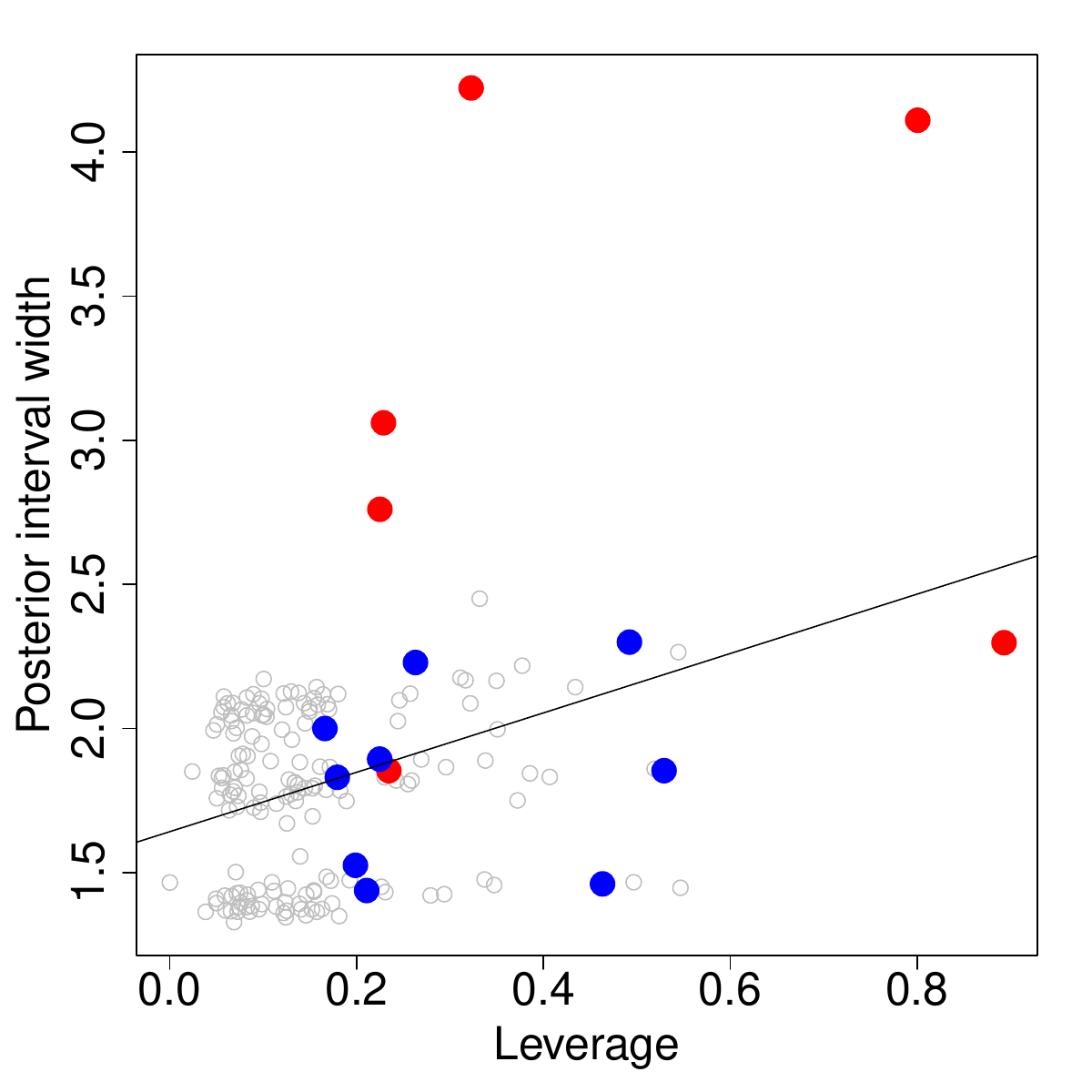}
\end{subfigure}
\begin{subfigure}[t]{.45\textwidth}
    \caption{ES}\label{pred_2020_ES_lev}
	\includegraphics[scale=0.35]{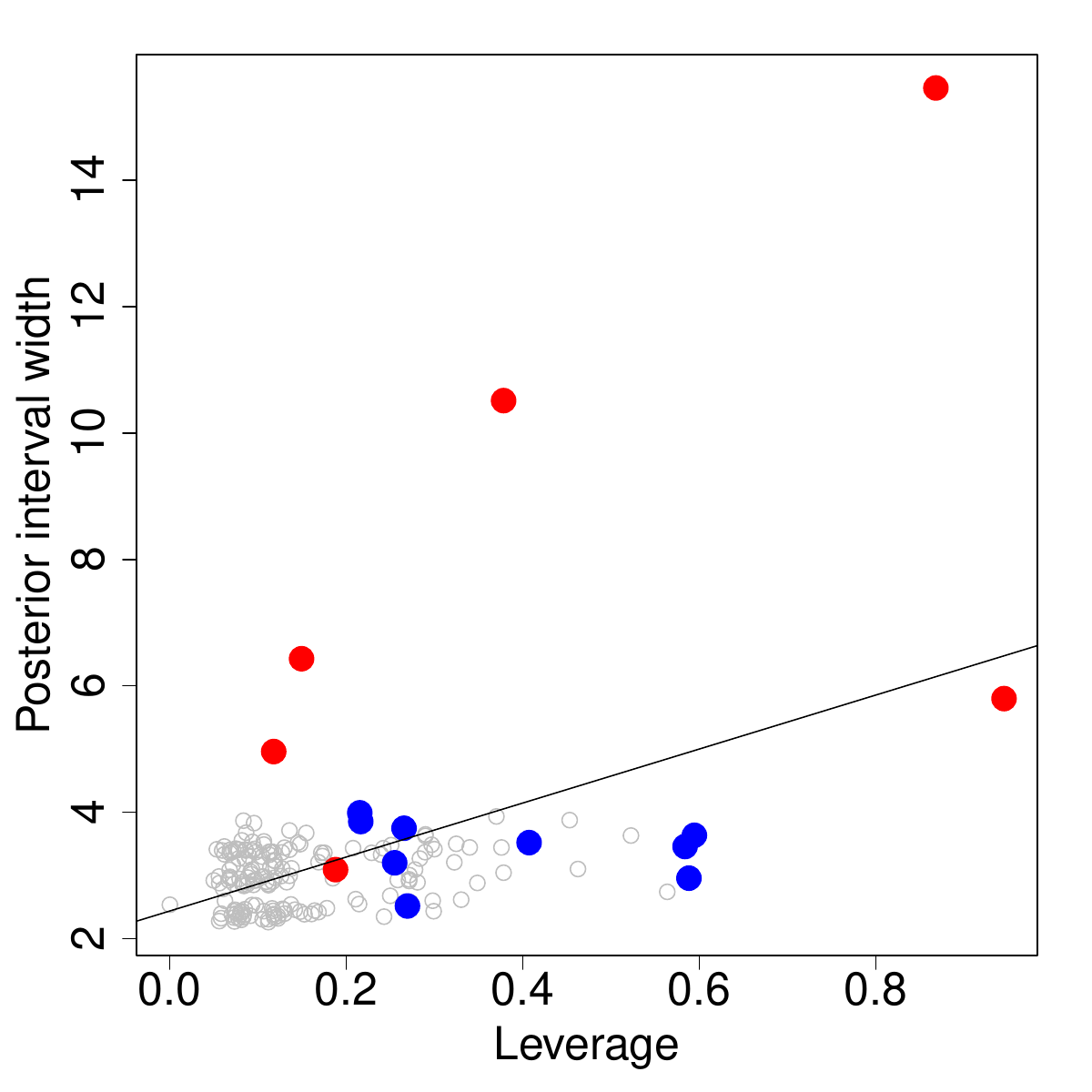}
\end{subfigure}
~
\begin{subfigure}[t]{.45\textwidth}
    \caption{FR}\label{pred_2020_FR_lev}
	\includegraphics[scale=0.35]{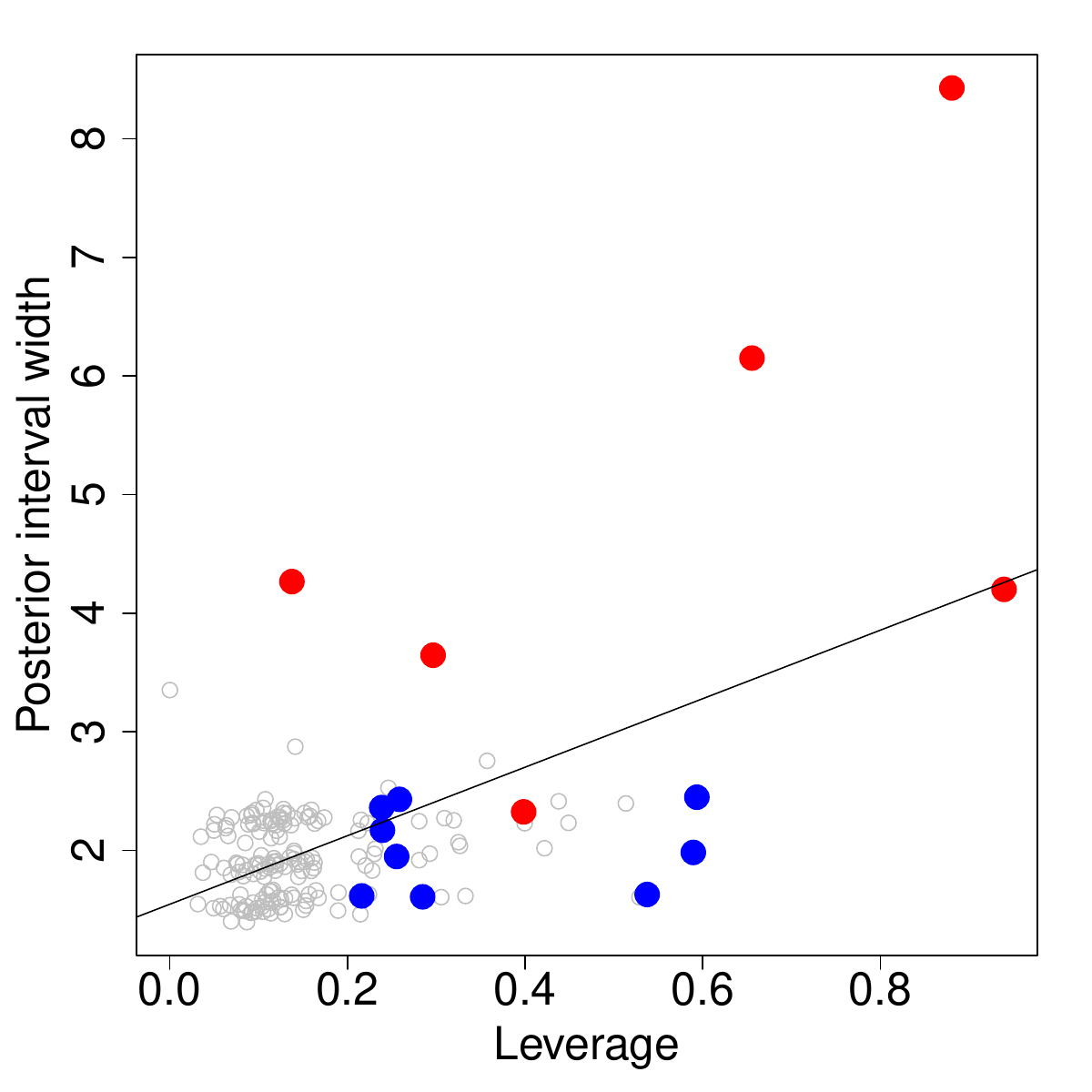}
\end{subfigure}
~
\begin{subfigure}[t]{.45\textwidth}
    \caption{IT}\label{pred_2020_IT_lev}
	\includegraphics[scale=0.35]{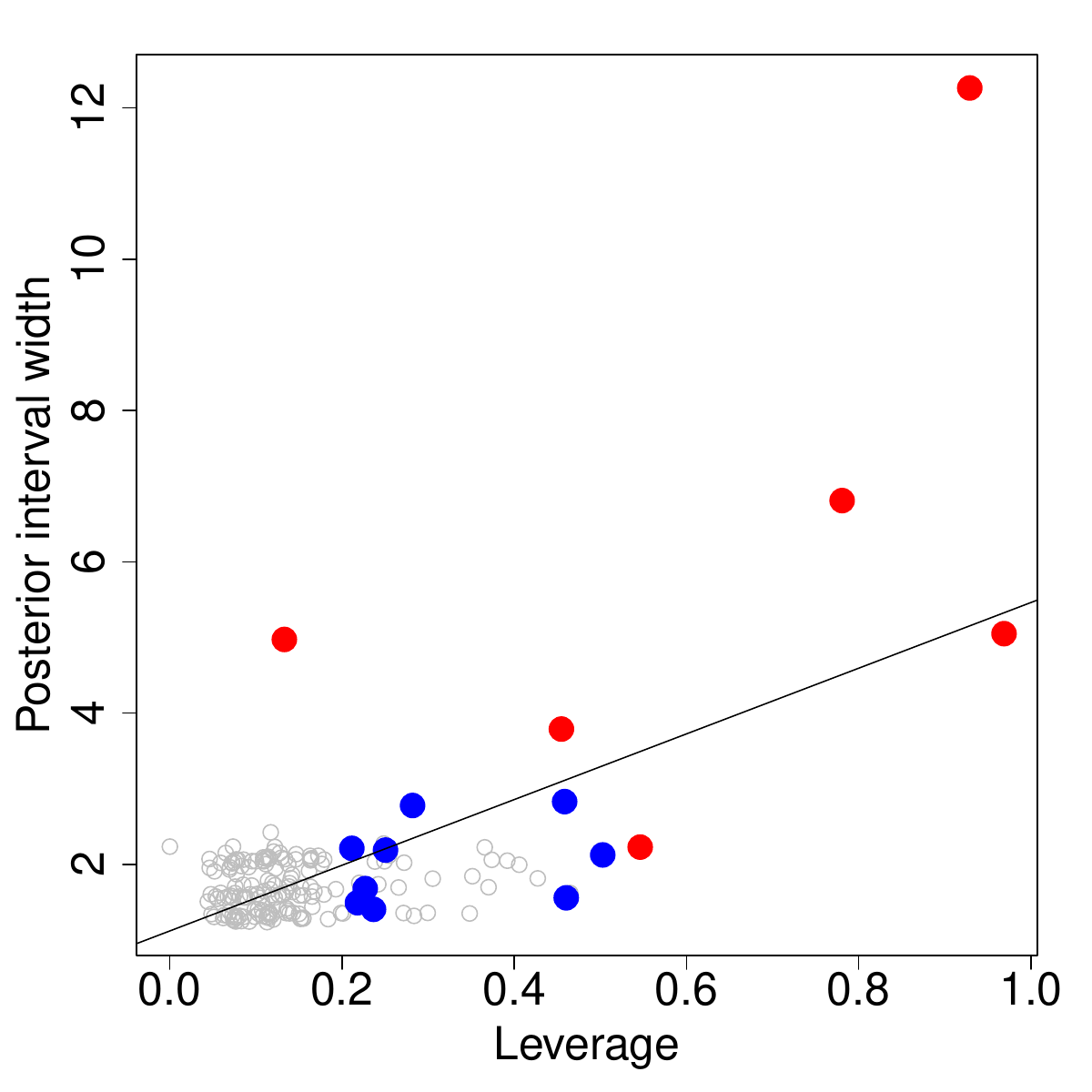}
\end{subfigure}
	\caption*{\footnotesize\textit{Notes}: The scatterplot shows the posterior interval width (defined as the difference between the $5^{\text{th}}$ and $95^{\text{th}}$ percentiles of the posterior distribution) against leverage. The red dots refer to the pandemic observations and the blue dots refer to the global financial crisis (i.e., 2008M06 to 2009M03) while the solid black line is the line of best fit.}
\end{figure}

We empirically demonstrate this effect in 
\autoref{fig: leverage}. 
This plots the posterior interval width (defined as the difference between $5^{\text{th}}$ and $95^{\text{th}}$ percentiles of the posterior distribution) of latent GDP against the leverage for all $T$ values of each. The leverage values $l_{t}~(t=1,\dots,T)$ are the diagonal elements of the projection matrix $\bm X (\bm X' \bm X)^{-1} \bm X'$. Since parts of $\bm X$ are unobserved in our MF model, we compute the leverage scores for each draw of the latent quantities and then compute the posterior mean of the scores.\footnote{Plugging in the posterior mean of the latent states to compute the leverage values leads to results which are qualitatively very similar.}

The figure clearly shows that there exists a positive relationship between leverage and posterior uncertainty. This implies that if we observe outlying observations (such as during the pandemic), our model yields wider credible sets. This feature is especially pronounced during the pandemic (the red dots in the figure) but we also find substantial evidence of a similar relationship during the global financial crisis (the blue dots). We conjecture that this is the main mechanism at work that yields improved predictive distributions through sharp and timely increases in the predictive variance.

\begin{figure}[!ht]
\caption{Log predictive likelihoods during the first two quarters of the pandemic for different numbers of trees.}\label{lps_treenum}
\centering
	\includegraphics[width=0.9\textwidth]{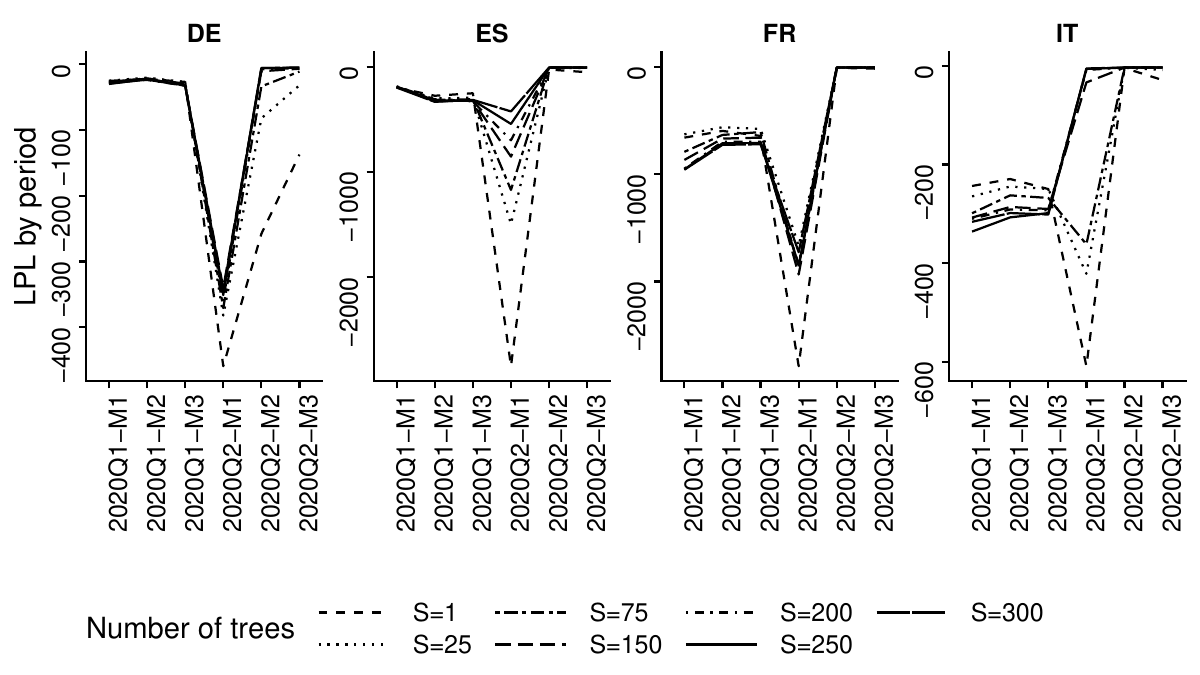}
	\caption*{\footnotesize\textit{Notes}: Log predictive likelihood (LPL) by period for 2020Q1 and 2020Q2, with M1, M2 and M3 indicating the month during the quarter when the nowcast is produced. $S$ refers to the number of trees used for BART, with $S=250$ being the benchmark specification used for our empirical work.}
\end{figure}

Adding more trees (i.e., increasing $S$) to capture the pandemic-related outliers will further increase the predictive variance relative to a model which sets $S$ small. To support this statement consider \autoref{lps_treenum} which shows log predictive likelihoods (LPLs) by period for the first two quarters of 2020, with nowcasts updated each month for all four countries. The lines in this figure vary in the number of trees used, ranging from a single tree to $250$ trees. While we observe several differences, these tend to become smaller the more trees we include, with $250$ trees indicating superior performance in almost all cases. However, once we get to $150$ trees, further forecast improvements tend to be small.

Another reason for its good performance is that BART is able to detect these tail events not only through ``remembering'' partitions of the  covariate space during certain periods such as the global financial crisis but also through recognizing that specific combinations of the covariates have not been observed in the past and deserve creating a new tree (or terminal node within a single tree). Again, such a new tree (or terminal node) will likely be equipped with relatively few observations and little information on the conditional mean function. This, in turn, will again (correctly) lead to inflated posterior intervals.

\subsubsection{Assessing Model Calibration using Probability Integral Transforms}
The preceding sub-section compared the relative performance of the MF-BAVART to the MF-VAR, but did not present any evidence on the nowcast performance of either in an absolute sense. In the Online Appendix, we provide graphs of the nowcasts of both approaches plotted against realized GDP growth for the four countries and three monthly nowcasts within each quarter. An examination of them indicates that the MF-BAVART's nowcasts are better calibrated, particularly for Spain. 

\begin{table*}[t]
\caption{Summary statistics of transformed PITs.}\vspace*{-1.5em}
\begin{center}
\begin{scriptsize}
\begin{threeparttable}
\begin{tabular*}{\textwidth}{@{\extracolsep{\fill}} ld{2.2}d{2.2}d{2.2}d{2.2}d{2.2}d{2.2}d{2.2}d{2.2}d{2.2}d{2.2}d{2.2}d{2.2}}
  \toprule
   & \multicolumn{6}{c}{\textbf{Through 2019Q4}} & \multicolumn{6}{c}{\textbf{Through 2020Q2}}\\
   \cmidrule(lr){2-7}\cmidrule(lr){8-13}
   & \multicolumn{3}{c}{\textbf{MF-BAVART}} & \multicolumn{3}{c}{\textbf{MF-VAR}} & \multicolumn{3}{c}{\textbf{MF-BAVART}} & \multicolumn{3}{c}{\textbf{MF-VAR}}\\
   \cmidrule(lr){2-4}\cmidrule(lr){5-7}\cmidrule(lr){8-10}\cmidrule(lr){11-13}
 \textbf{Mth./Qt.} & \multicolumn{1}{c}{1} & \multicolumn{1}{c}{2} & \multicolumn{1}{c}{3} & \multicolumn{1}{c}{1} & \multicolumn{1}{c}{2} & \multicolumn{1}{c}{3} & \multicolumn{1}{c}{1} & \multicolumn{1}{c}{2} & \multicolumn{1}{c}{3} & \multicolumn{1}{c}{1} & \multicolumn{1}{c}{2} & \multicolumn{1}{c}{3} \\
  \midrule
  \textbf{DE} & & & & & & & & & & & & \\
$\mu$ & 0.44 & 0.37 & 0.45 & 0.83 & 0.91 & 0.87 & -0.48 & 0.06 & 0.12 & -0.27 & 0.43 & 0.39 \\ 
  $\sigma^2$ & 3.59 & 3.52 & 3.12 & 6.97 & 6.85 & 4.14 & 22.97 & 5.20 & 5.23 & 35.58 & 10.87 & 8.79 \\ 
  AR(1) & -0.33 & -0.32 & -0.37 & -0.29 & -0.15 & -0.18 & 0.83 & -0.07 & -0.12 & 0.60 & 0.08 & 0.04 \\ 
  \midrule
  \textbf{ES} & & & & & & & & & & & & \\
  $\mu$ & -0.25 & -0.66 & -0.56 & -0.29 & -0.73 & -0.58 & -1.45 & -1.45 & -1.31 & -1.55 & -1.97 & -1.72 \\ 
  $\sigma^2$ & 2.33 & 2.52 & 2.77 & 10.56 & 8.60 & 7.06 & 30.64 & 19.79 & 19.77 & 41.35 & 42.65 & 36.42 \\ 
  AR(1) & 0.71 & 0.51 & 0.56 & 0.72 & 0.65 & 0.58 & 1.24 & 0.29 & 0.24 & 0.69 & 0.52 & 0.45 \\ 
  \midrule
  \textbf{FR} & & & & & & & & & & & & \\
  $\mu$ & 0.34 & 0.35 & 0.33 & 0.49 & 0.69 & 0.55 & -0.70 & -0.76 & -0.3 & -0.57 & -0.10 & -0.16 \\ 
  $\sigma^2$ & 3.34 & 3.02 & 3.50 & 5.53 & 5.07 & 4.54 & 24.37 & 40.29 & 14.73 & 26.77 & 18.47 & 16.35 \\ 
  AR(1) & 0.17 & 0.15 & -0.06 & 0.19 & 0.12 & 0.09 & 0.93 & 0.22 & 0.28 & 0.83 & 0.46 & 0.37 \\ 
  \midrule
  \textbf{IT} & & & & & & & & & & & & \\
  $\mu$ & -0.10 & -0.05 & 0.01 & -0.05 & 0.06 & 0.18 & -0.96 & -0.79 & -0.74 & -1.45 & -0.99 & -0.84 \\ 
  $\sigma^2$ & 1.54 & 1.38 & 1.17 & 5.67 & 4.45 & 3.09 & 19.58 & 17.51 & 17.38 & 47.28 & 30.66 & 33.35 \\ 
  AR(1) & 0.60 & 0.46 & 0.32 & 0.58 & 0.40 & 0.30 & 0.39 & 0.23 & 0.21 & 0.51 & 0.35 & 0.20 \\ 
  \bottomrule
\end{tabular*}
\begin{tablenotes}[para,flushleft]
\scriptsize{\textit{Notes}: Mth./Qt. denotes which month within the quarter the nowcast was made. $\mu$, $\sigma^2$ and AR(1) denote the sample mean, variance and AR(1) coefficient of the transformed probability intergral transforms (PITs).}
\end{tablenotes}
\end{threeparttable}
\end{scriptsize}
\end{center}
\label{PITs2019}
\end{table*}

In this sub-section, we investigate this issue more formally using Probability Integral Transforms (PITs). In particular, we follow a common practice \citep[e.g.,][]{Clark2011} and produce PITs for our nowcasts and transform them using the inverse of the c.d.f. of a standard Gaussian distribution. We denote these transformed PITs as $r_{t}$ for the time of our nowcast evaluation period. Perfectly calibrated nowcasts should lead to $r_t$ having mean zero, variance one and being uncorrelated over time. We calculate the sample mean (labeled $\mu$ in the tables), variance (labeled $\sigma^2$) and estimated autoregressive coefficient (labeled AR(1)). Table \ref{PITs2019} displays these summary statistics for the sample through 2019Q4 and the full sample, respectively. 

Beginning with the linear MF-VAR, note that even in the pre-pandemic sample, there is some evidence of poor calibration. For the sample mean, the point estimates are consistently well away from zero. The sample variances are often substantially higher than one indicating the predictive variance of the linear model is too small. There is sometimes evidence of autocorrelation in $r_{t}$. When we move to the full sample, these problems get much worse, particularly for the sample variance of $r_t$ which now becomes very large. 

If we turn to the MF-BAVART for the sample ranging through 2020Q2, it can be seen that the nowcasts are better calibrated. Even for the full sample, the sample mean of $r_t$ is often close to zero. The estimated AR(1) coefficient  often indicates autocorrelation in the first month per quarter, while second and third month nowcasts indicate more favorable calibration. It is the case that the sample variance of $r_t$ is still too high, but to a much lesser extent than for the MF-VAR. Thus, use of the MF-BAVART has gone a large way towards improving the calibration problems of the MF-VAR, even if it has not completely fixed them. 

\subsubsection{A Closer Look at the Pandemic Period}
In this sub-section, we provide additional insight as to how MF-BAVART is nowcasting in the two pandemic quarters. In addition to taking a closer look at the densities produced by the MF-BAVART and MF-VAR models (labeled ``Benchmark'' models in the following figures) of the preceding sub-sections, we also produce results for an eight dimensional MF-BAVART which adds the two pandemic variables (see Sub-section \ref{data}, this is labeled the ``Pandemic'' model in the figures).\footnote{Since the pandemic variables are non-zero for an extremely short time and feature rapid shifts, the MF-VAR was not capable of handling them without substantial prior tuning and hence we do not include results from the MF-VAR using the pandemic variables.}

\begin{table*}[t]
\caption{Log predictive likelihoods for MF-BAVART for each month within the first two quarters of 2020.}\vspace*{-1.5em}
\begin{center}
\begin{scriptsize}
\begin{threeparttable}
\begin{tabular*}{\textwidth}{@{\extracolsep{\fill}} ld{3.2}d{3.2}d{3.2}d{4.2}d{1.2}d{1.2}d{3.2}d{2.2}d{3.2}d{1.2}d{1.2}d{1.2}}
  \toprule
  & \multicolumn{6}{c}{\textbf{Benchmark}} & \multicolumn{6}{c}{\textbf{Pandemic}}\\
   \cmidrule(lr){2-7}\cmidrule(lr){8-13}
  & \multicolumn{3}{c}{\textbf{2020Q1}} & \multicolumn{3}{c}{\textbf{2020Q2}} & \multicolumn{3}{c}{\textbf{2020Q1}} & \multicolumn{3}{c}{\textbf{2020Q2}} \\
  \cmidrule(lr){2-4}\cmidrule(lr){5-7}\cmidrule(lr){8-10}\cmidrule(lr){11-13}
 \textbf{Mth./Qt.} & \multicolumn{1}{c}{1} & \multicolumn{1}{c}{2} & \multicolumn{1}{c}{3} & \multicolumn{1}{c}{1} & \multicolumn{1}{c}{2} & \multicolumn{1}{c}{3} & \multicolumn{1}{c}{1} & \multicolumn{1}{c}{2} & \multicolumn{1}{c}{3} & \multicolumn{1}{c}{1} & \multicolumn{1}{c}{2} & \multicolumn{1}{c}{3}\\
  \midrule
  \textbf{DE} & -29.6 & -22.6 & -32.2 & -349.8 & -6.3 & -4.9 & -10.9 & -2.3 & -12.4 & -4.2 & -3.8 & -3.5 \\ 
  \textbf{ES} & -190.4 & -337.3 & -321.6 & -538.3 & -6.7 & -5.4 & -77.8 & -94.3 & -3.3 & -6.7 & -4.8 & -4.2 \\ 
  \textbf{FR} & -925.4 & -721.2 & -696.6 & -1918.6 & -4.1 & -4.1 & -346.4 & -15.3 & -107.4 & -9.1 & -4.6 & -4.0 \\ 
  \textbf{IT} & -319.4 & -291.8 & -307.5 & -6.3 & -3.8 & -3.6 & -92.5 & -5.7 & -29.6 & -4.6 & -3.7 & -4.0 \\ 
   \bottomrule
\end{tabular*}
\begin{tablenotes}[para,flushleft]
\scriptsize{\textit{Notes}: ``Benchmark'' refers to the information set described above in the context of our main nowcast evaluation. ``Pandemic'' includes further variables measuring infection rates and Google mobility trends. Mth./Qt. denotes which month within the quarter the nowcast was made. }
\end{tablenotes}
\end{threeparttable}
\end{scriptsize}
\end{center}
\label{tab:lps2020}
\end{table*}

In Sub-section \ref{compare}, we showed that (with one exception) the benchmark MF-BAVART  produces greatly superior predictive likelihoods relative to the MF-VAR during the six pandemic months. Table \ref{tab:lps2020} offers additional insight by comparing log predictive likelihoods for the benchmark and pandemic versions of the MF-BAVART model. It can be seen that adding the pandemic variables often improves the nowcasting performance of the MF-BAVART. There are several exceptions to this, but in the crucial pandemic months of March, April and May the inclusion of pandemic variables is beneficial in all four countries. Thus, despite the very short time series and the very simple way these variables are included in the model, BART finds a way to usefully incorporate the information they contain.

One interesting result is that both time series even improve nowcasts produced in January and February 2020. Inspection of the pandemic time series reveals that all countries have reported the first infections already at the end of January. Since we assume that the nowcasts are always produced towards the end of each month, BART is essentially using this increase from zero to very few infections in combination with deteriorating values in the PMI, car registrations and the ESI. As we will show below, this does not signal a downturn in GDP but merely reflects the fact that uncertainty surrounding the predictive distribution is increasing (which also happens technically because the size of the model is increased as well). And this is beneficial for nowcasting.

\begin{figure}[!ht]
\caption{Predictive densities for the first two quarters of 2020 across countries.}
\centering
\begin{subfigure}[t]{\textwidth}
    \caption{DE}\label{pred_2020_GE}
	\includegraphics[width=\textwidth]{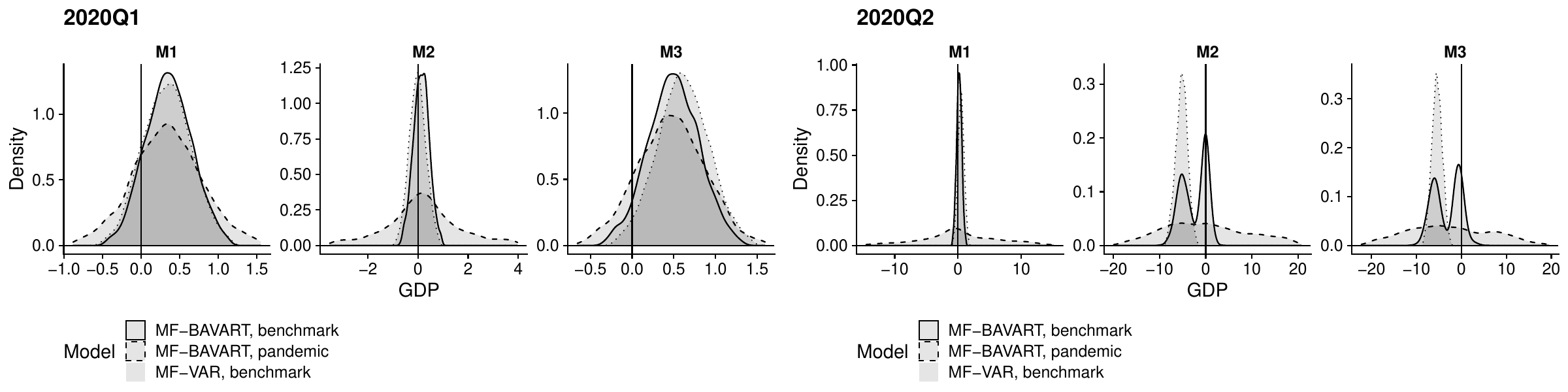}
\end{subfigure}
~
\begin{subfigure}[t]{\textwidth}
    \caption{ES}\label{pred_2020_ES}
	\includegraphics[width=\textwidth]{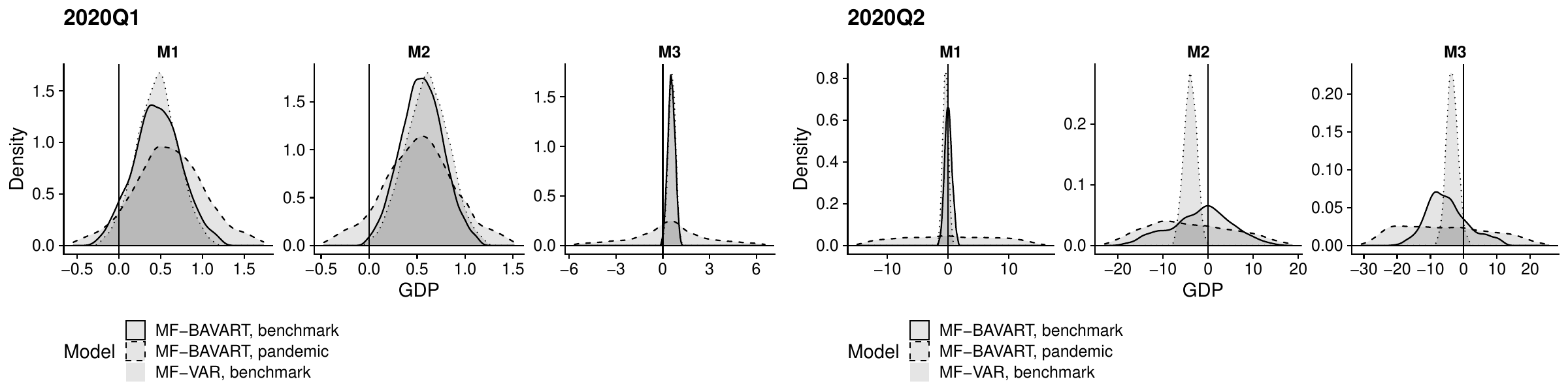}
\end{subfigure}
~
\begin{subfigure}[t]{\textwidth}
    \caption{FR}\label{pred_2020_FR}
	\includegraphics[width=\textwidth]{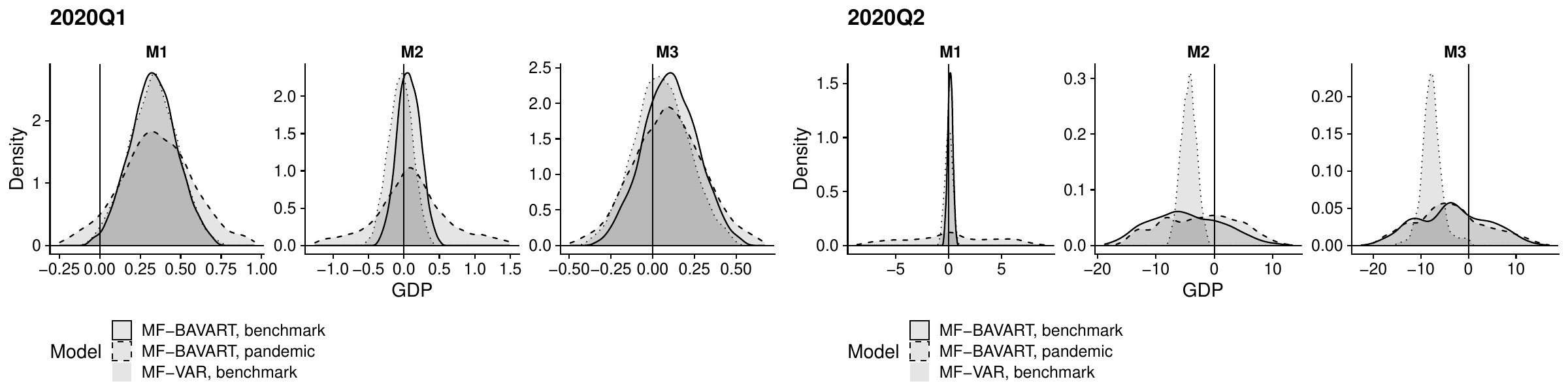}
\end{subfigure}
~
\begin{subfigure}[t]{\textwidth}
    \caption{IT}\label{pred_2020_IT}
	\includegraphics[width=\textwidth]{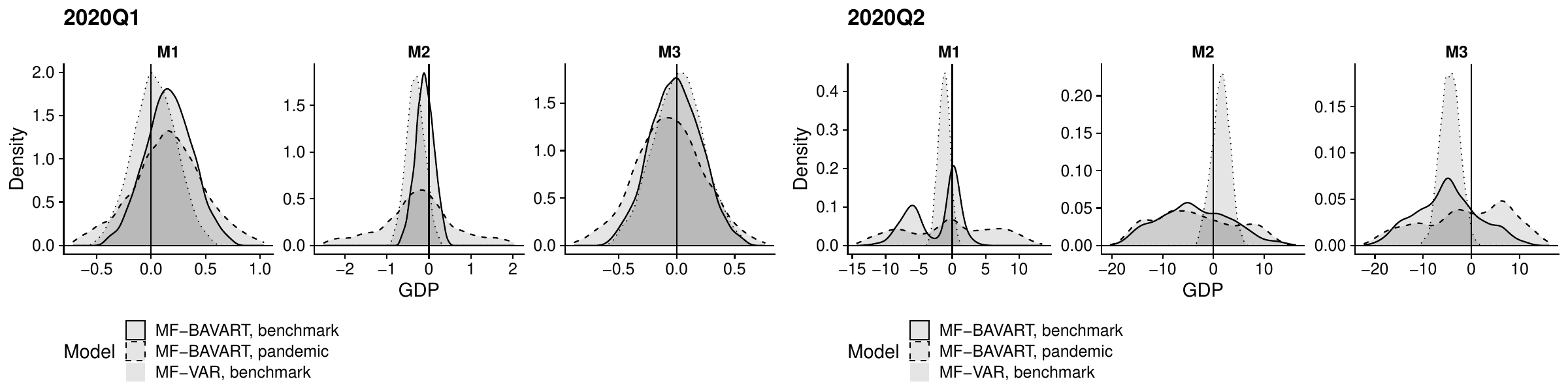}
\end{subfigure}
\label{preds}
\caption*{\footnotesize\textit{Notes}: ``Benchmark'' refers to the information set described above in the context of our main nowcast evaluation. ``Pandemic'' includes further variables measuring infection rates and Google mobility trends. The black vertical line marks zero.}
\end{figure}

Figure \ref{preds} plots the predictive densities themselves for the three models for the first six months of 2020. The key general finding is that, as expected, MF-BAVART is much more flexible than the MF-VAR. Particularly in 2020Q2, the predictive densities it produces tend to be much more dispersed, feature fatter tails, are often asymmetric and even sometimes multi-modal. This contrasts with the MF-VAR where the predictive densities tend to be closer to Gaussian densities. 

In light of the recent interest of macroeconomics in models involving asymmetries and multi-modalities \citep[see, e.g.,][]{adrianetal,vulnerable} this feature of the MF-BAVART is particularly attractive and is the source of the improvements in nowcast performance during the pandemic. If we compare the benchmark and pandemic versions of the non-parametric model, a clear pattern emerges. As the pandemic hits, the predictive densities for the pandemic model quickly become much more dispersed than those of the benchmark model. The latter, in turn, are more dispersed than those of the linear model. BART is deciding to incorporate information in these short time series in such a way that the predictive densities reflect the increased uncertainty.  This is due to the fact that inclusion of these two additional very short time series is akin to adding a time-varying intercept to the model with a rather large state innovation variance from January 2020 onward. And this random intercept term scales up the variance of the predictive distribution and thus improves density nowcasts (especially in the beginning of the pandemic period).

\section{Summary and Conclusions}\label{sec:summary}
MF-VARs have been a standard tool for producing timely, high frequency nowcasts of low frequency variables for several years. With the arrival of the COVID-19 pandemic of 2020 the need for such nowcasts has become even more acute. However, conventional linear MF-VARs nowcast poorly during the pandemic due to their inability to effectively deal with the extreme observations that have occurred. In this paper, we have developed the MF-BAVART which is a non-parametric model using additive regression trees. MF-BAVART can be cast as a non-linear state space model. We develop an approximate MCMC algorithm where the parameters defining the conditional mean of the VAR are drawn using a standard BART algorithm and, conditional on these, the states are drawn using a linear approximation. This linear approximation is taken from the machine learning literature on black-box models and we use simulations to show that it also works well for a DGP that closely matches the evolution of GDP during the pandemic.

Our nowcasting exercise, involving four major euro area countries, shows that MF-BAVART, with few exceptions, performs better than the linear MF-VAR at all times in our sample, with major nowcasting benefits during the pandemic. We show how and why this occurs by providing a detailed comparison of nowcast densities in the first six months of 2020. 

The techniques outlined in this paper have a wide range of potential applications, as they can be applied to any non-linear and non-parametric learner commonly used in the literature. Our focus on using BART is motivated by its strong performance in various applications \citep[see, e.g.,][]{bleich2014variable, Linero2018,kapelner2015prediction} as well as its flexibility in handling outliers. As a fruitful avenue for further research one could assess how different learners perform and then combine them using Bayesian model averaging techniques.

\small{\setstretch{0.85}
\addcontentsline{toc}{section}{References}
\bibliographystyle{custom.bst}
\bibliography{MF_BART_Biblio}}\normalsize\clearpage

\begin{center}\LARGE\textbf{Appendix}\end{center}
\begin{appendices}
\setcounter{equation}{0}
\renewcommand\theequation{A.\arabic{equation}}
\section{Equation-by-Equation Estimation of the VAR} \label{sec: AppMCMC}
In this Appendix we show how to rewrite the VAR as a system of unrelated regression models. This approach has the advantage that the computational burden is drastically reduced since we can perform equation-by-equation estimation.    Notice that the first equation of Eq. (\ref{nonVAR})  can be written as:
\begin{equation*}
y_{1t} = f_1 ( \bm X_t) + \eta_{1t}, \quad \eta_{1t} \sim \mathcal{N}(0, \sigma_1^2).
\end{equation*}
The second equation is given by:
\begin{equation*}
y_{2t} = f_2 ( \bm X_t) + q_{21} \eta_{1t} + \eta_{2t}, \quad \eta_{2t} \sim \mathcal{N}(0, \sigma_2^2).
\end{equation*}
In general, the $j^{\text{\text{th}}} > 1$ equation can be written as:
\begin{equation}
y_{jt} = f_j(\bm X_t) +\bm q_j' \bm Z_{jt} + \eta_{jt}, \quad \eta_{jt} \sim \mathcal{N}(0, \sigma_j^2). \label{eq: struceq-by-eq}
\end{equation}
This implies that, conditional on the shocks to the previous $j-1$ equations, the $j^{\text{th}}$ equation is a standard regression model that features a non-parametric part given by $f_j(\bm X_t)$ and a regression part $\bm q_j' \bm Z_{jt}$ with $\bm q_j =(  q_{j1}, \dots, q_{jj-1})'$ and $\bm Z_{jt} = (\eta_{1t}, \dots, \eta_{j-1t})'$.  The $(j-1)$-dimensional vector $\bm q_j$ stores the first $j-1$ elements of the  $j^{\text{th}}$ row of $\bm Q$. These equations are conditionally independent and standard MCMC techniques can be readily applied. 

Alternative algorithms replace the shocks with the contemporaneous values of $\bm y_t$. This introduces order dependence which we avoid by conditioning on the shocks. Thus, we are using a standard sampling algorithm that is commonly used to sample from the multivariate Gaussian \citep{Carriero2019}.

\clearpage
\begin{center}\LARGE\textbf{Online Appendix}\end{center}
\setcounter{figure}{0}
\renewcommand\thefigure{A.\arabic{figure}}
\renewcommand\thesection{A}
\section{Additional Empirical Results}
\subsection{Log Predictive Scores Over Time}
\begin{figure}
\caption{Log predictive scores for MF-BAVART (solid line) relative to MF-VAR (dashed).}
\centering
\begin{subfigure}[t]{\textwidth}
    \caption{DE}\label{lps_GE}
	\includegraphics[width=0.88\textwidth]{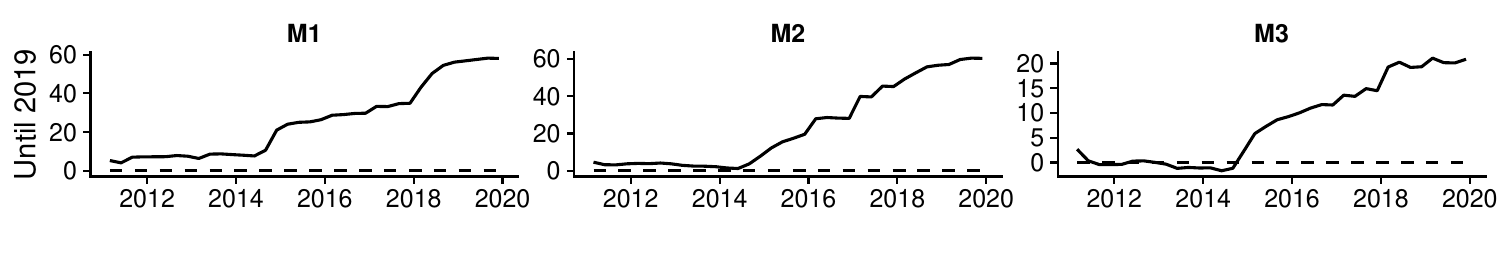}
	\includegraphics[width=0.88\textwidth]{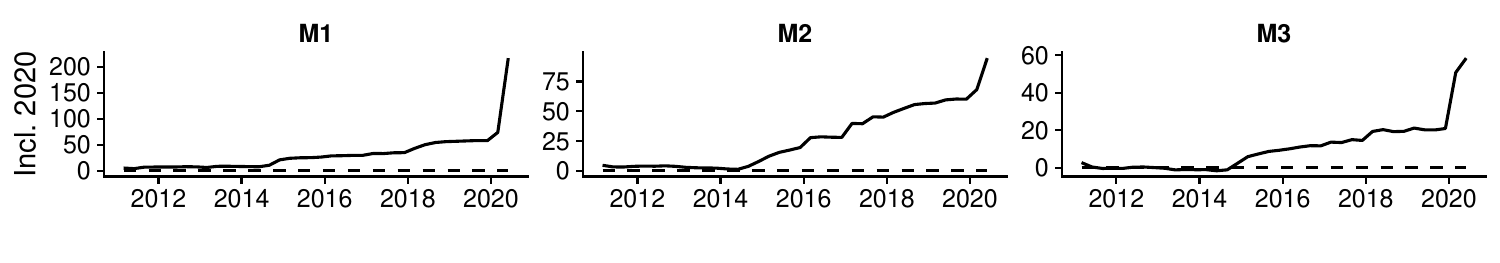}
\end{subfigure}
~
\begin{subfigure}[t]{\textwidth}
    \caption{ES}\label{lps_SP}
	\includegraphics[width=0.88\textwidth]{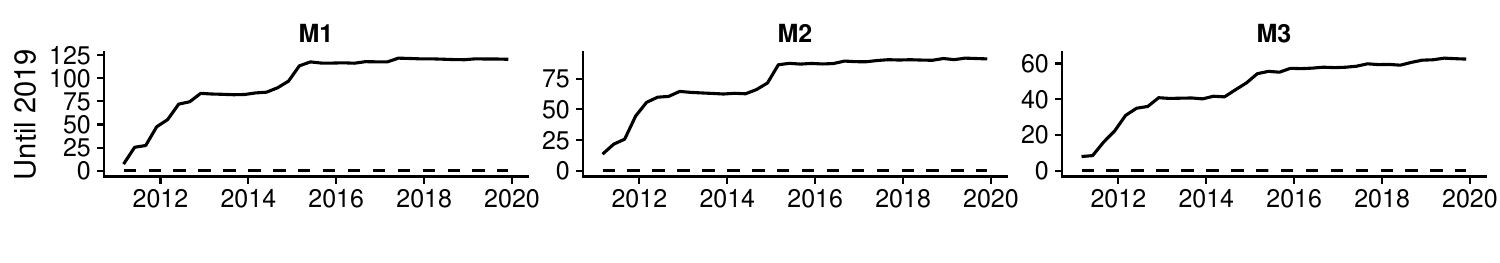}
	\includegraphics[width=0.88\textwidth]{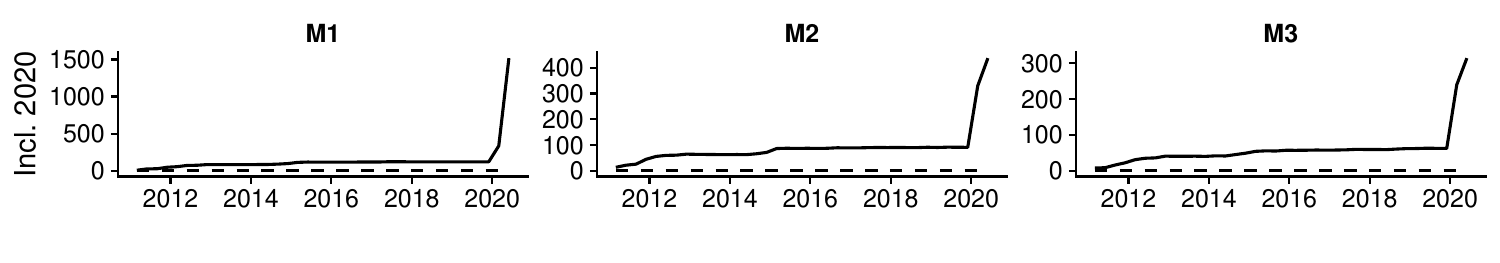}
\end{subfigure}
~
\begin{subfigure}[t]{\textwidth}
    \caption{FR}\label{lps_FR}
	\includegraphics[width=0.88\textwidth]{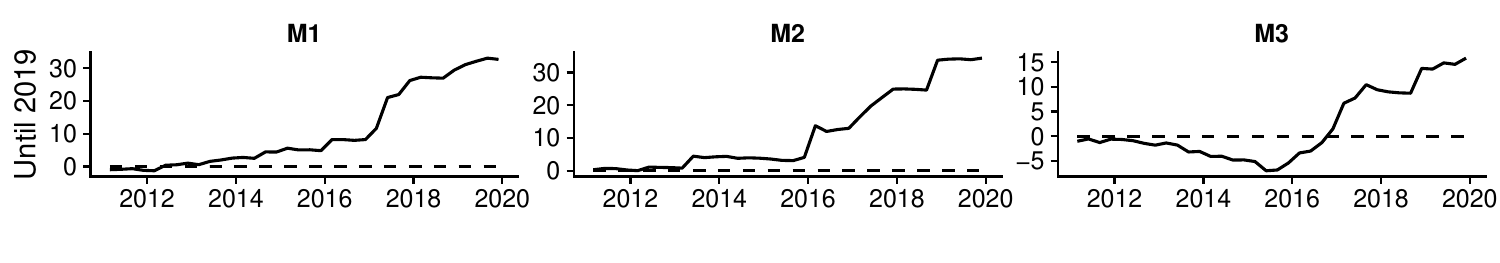}
	\includegraphics[width=0.88\textwidth]{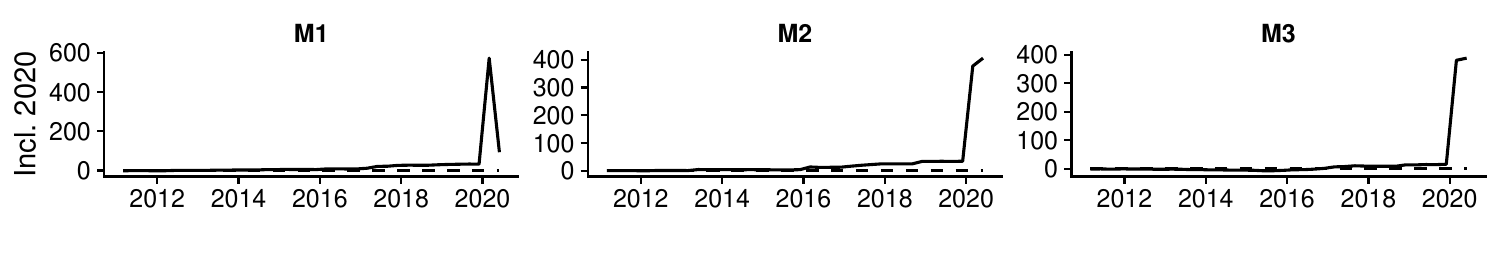}
\end{subfigure}
~
\begin{subfigure}[t]{\textwidth}
    \caption{IT}\label{lps_IT}
	\includegraphics[width=0.88\textwidth]{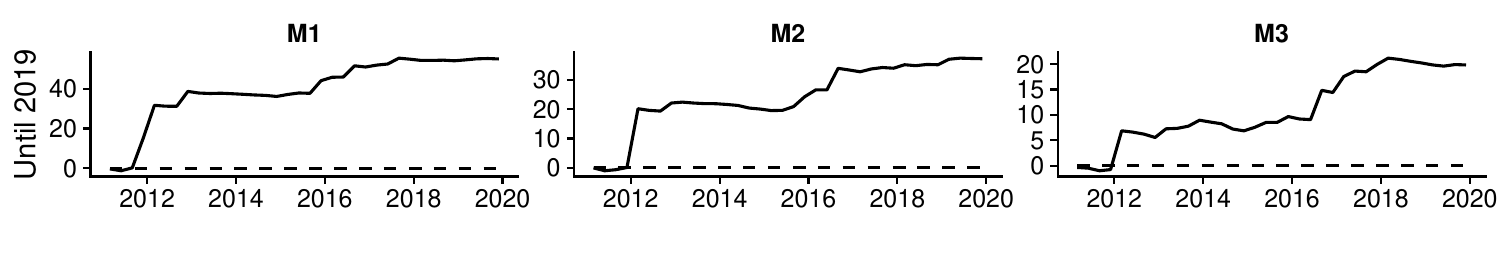}
	\includegraphics[width=0.88\textwidth]{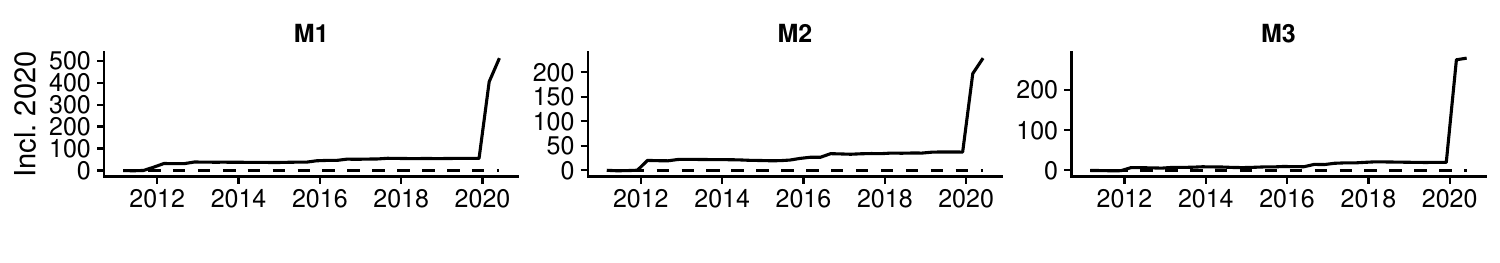}
\end{subfigure}
\caption*{\footnotesize\textit{Notes}: Samples ending in 2019Q4 are marked ``Until 2019,'' those ending in 2020Q2 are indicated as ``Incl. 2020.'' M1, M2 and M3 refer to the month during the quarter when the nowcast was produced.}
\end{figure}

The LPS is the preferred comparison metric for many Bayesians. To investigate patterns in it more deeply, consider Figures \ref{lps_GE}, \ref{lps_SP}, \ref{lps_FR} and \ref{lps_IT} which plot cumulative sums of log predictive likelihoods over time. There tends to be a pattern where the advantage in nowcast performance of the MF-BAVART relative to the MF-VAR steadily and gradually increases over time through 2019. Then in 2020, there tends to be a large jump in the LPS's of the MF-BAVART relative to the MF-VAR. There is only one exception to this: French nowcasts made by the MF-BAVART in the first month of each quarter show a large jump in 2020Q1, but then display a fall in 2020Q2. Thus, we are doing a poor nowcasting job in April 2020 for France relative to the linear model. But with this one exception, the MF-BAVART model is doing an excellent job of handling the pandemic.  

\subsection{Predictive Densities and Actual Realizations}
Here, we plot the nowcasts against the realizations. Our model produces monthly nowcasts of GDP growth which are converted into quarterly nowcasts to be comparable to the realization. To improve readability, we present results through 2019Q4 and through 2020Q2 as separate graphs. 
\begin{figure}[ht]
\caption{Predictive densities for Germany.}
\begin{subfigure}[t]{\textwidth}
	\caption{Until 2019}
    \includegraphics[width = 1\textwidth]{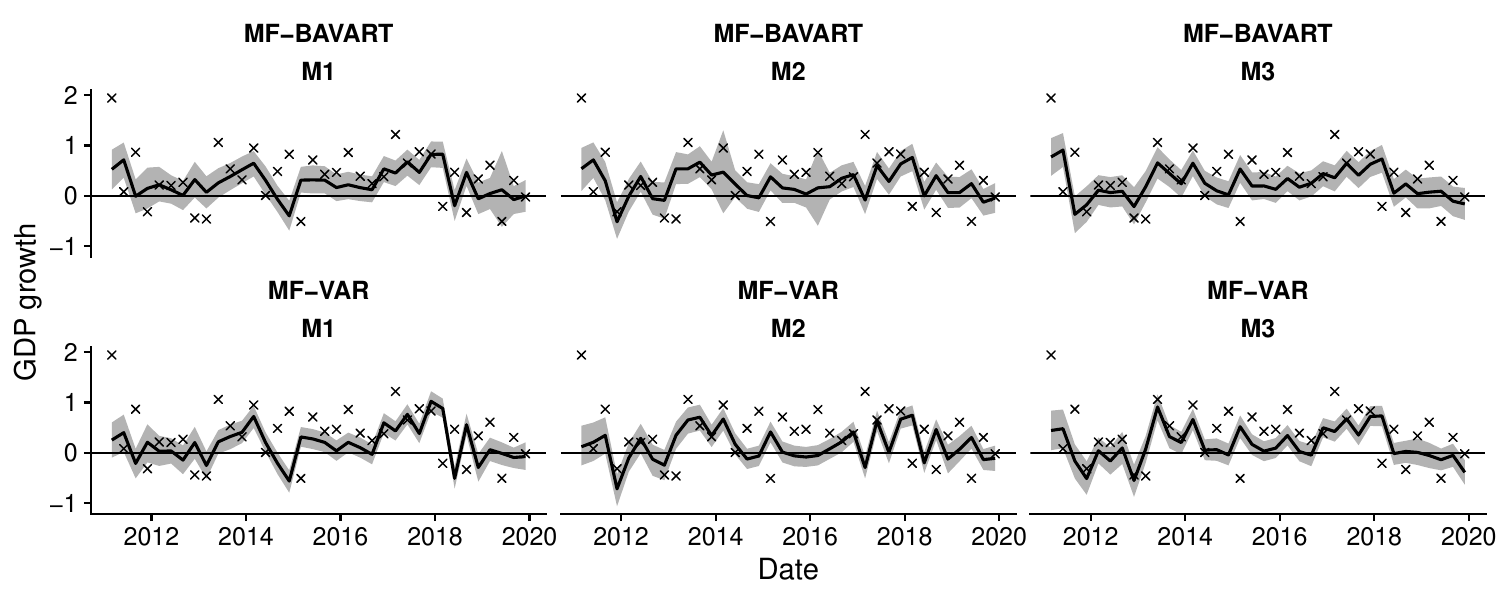}
\end{subfigure}
~
\begin{subfigure}[t]{\textwidth}
	\caption{Including the pandemic}
    \includegraphics[width = 1\textwidth]{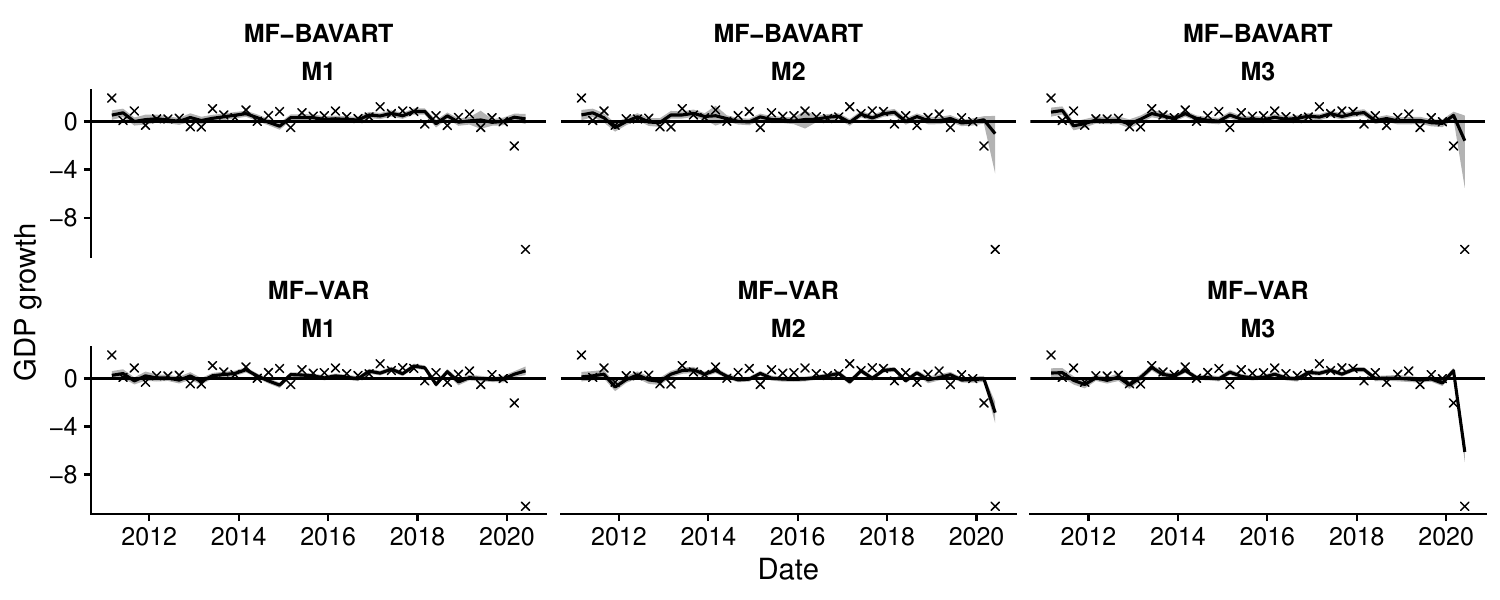}
\end{subfigure}
\caption*{\footnotesize\textit{Notes}: Columns are months per quarter in which the nowcast was produced. Realizations are marked as X's, and shown alongside the estimate for the posterior median and the 68 percent credible set.}
\end{figure}

\begin{figure}[ht]
\caption{Predictive densities for Spain.}
\begin{subfigure}[t]{\textwidth}
	\caption{Until 2019}
    \includegraphics[width = 1\textwidth]{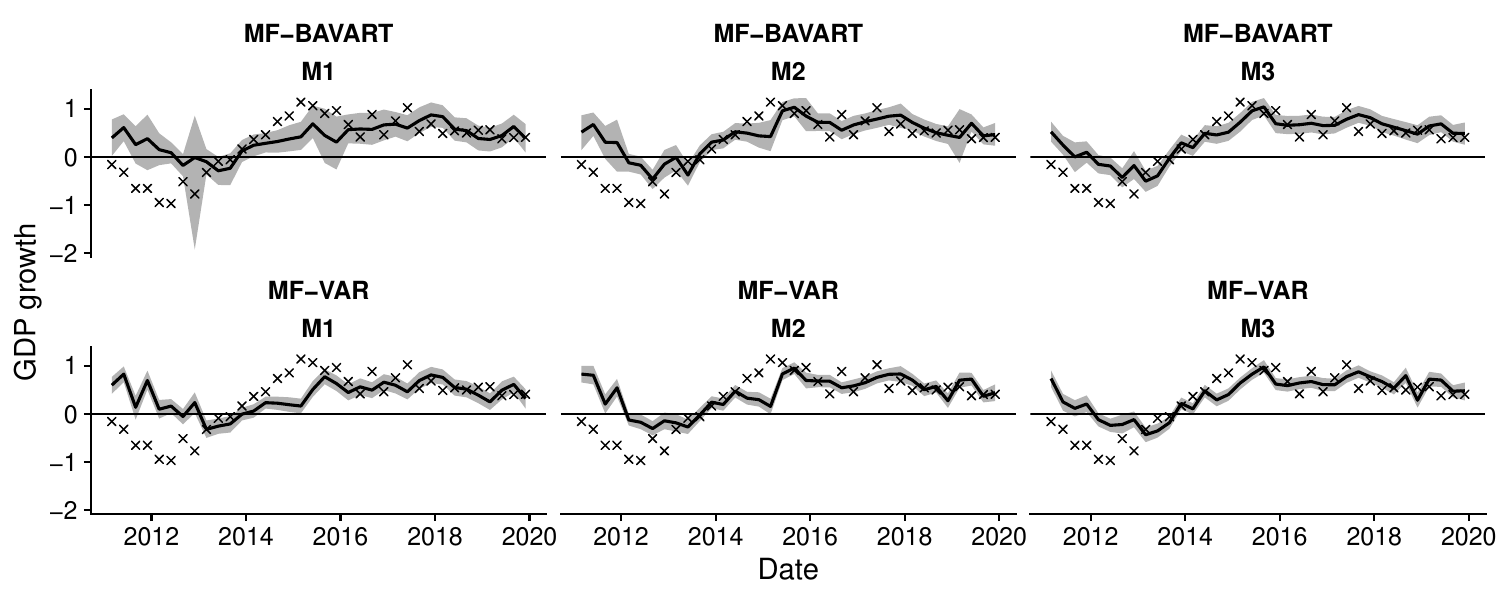}
\end{subfigure}
~
\begin{subfigure}[t]{\textwidth}
	\caption{Including the pandemic}
    \includegraphics[width = 1\textwidth]{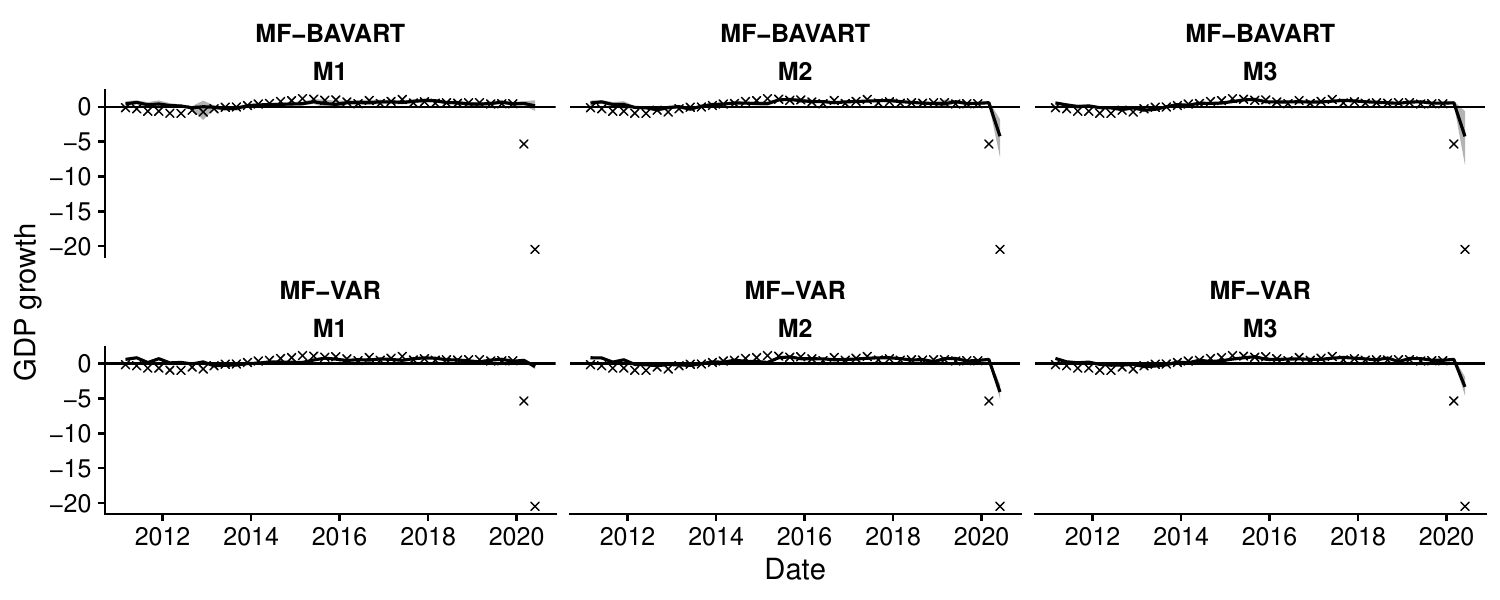}
\end{subfigure}
\caption*{\footnotesize\textit{Notes}: Columns are months per quarter in which the nowcast was produced. Realizations are marked as X's, and shown alongside the estimate for the posterior median and the 68 percent credible set.}
\end{figure}

\begin{figure}[ht]
\caption{Predictive densities for France.}
\begin{subfigure}[t]{\textwidth}
	\caption{Until 2019}
    \includegraphics[width = 1\textwidth]{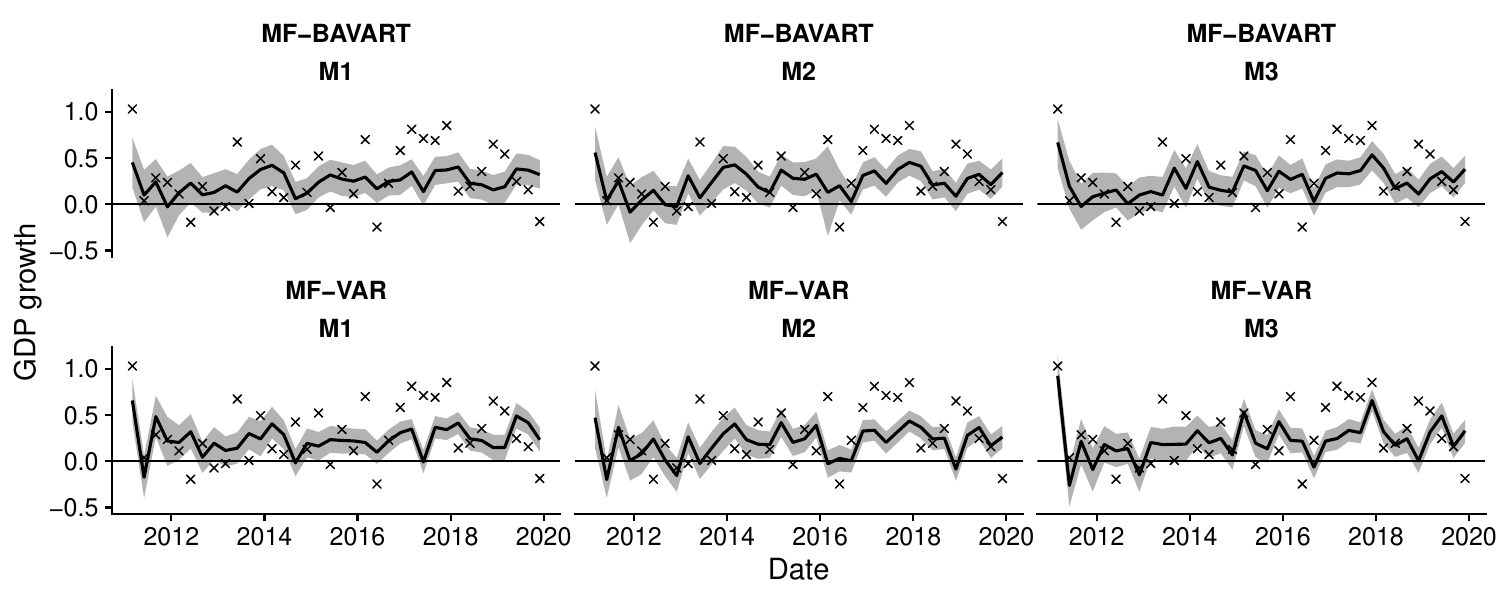}
\end{subfigure}
~
\begin{subfigure}[t]{\textwidth}
	\caption{Including the pandemic}
    \includegraphics[width = 1\textwidth]{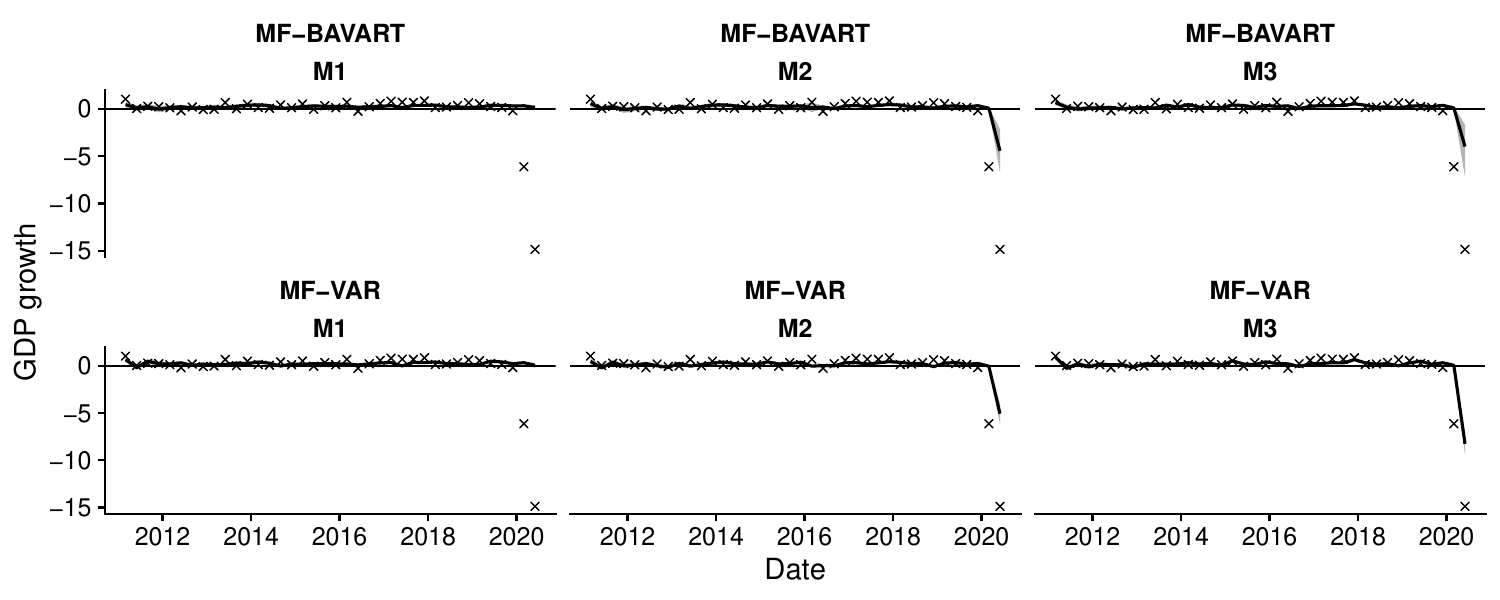}
\end{subfigure}
\caption*{\footnotesize\textit{Notes}: Columns are months per quarter in which the nowcast was produced. Realizations are marked as X's, and shown alongside the estimate for the posterior median and the 68 percent credible set.}
\end{figure}

\begin{figure}[ht]
\caption{Predictive densities for Italy.}
\begin{subfigure}[t]{\textwidth}
	\caption{Until 2019}
    \includegraphics[width = 1\textwidth]{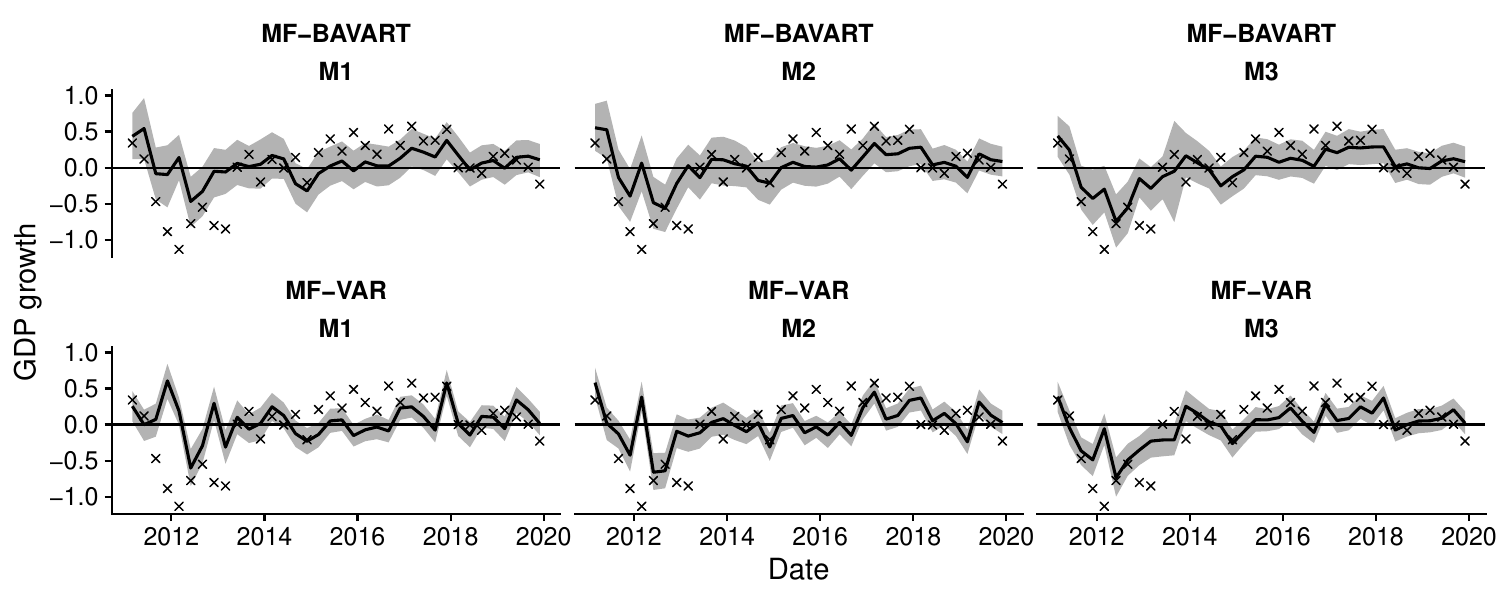}
\end{subfigure}
~
\begin{subfigure}[t]{\textwidth}
	\caption{Including the pandemic}
    \includegraphics[width = 1\textwidth]{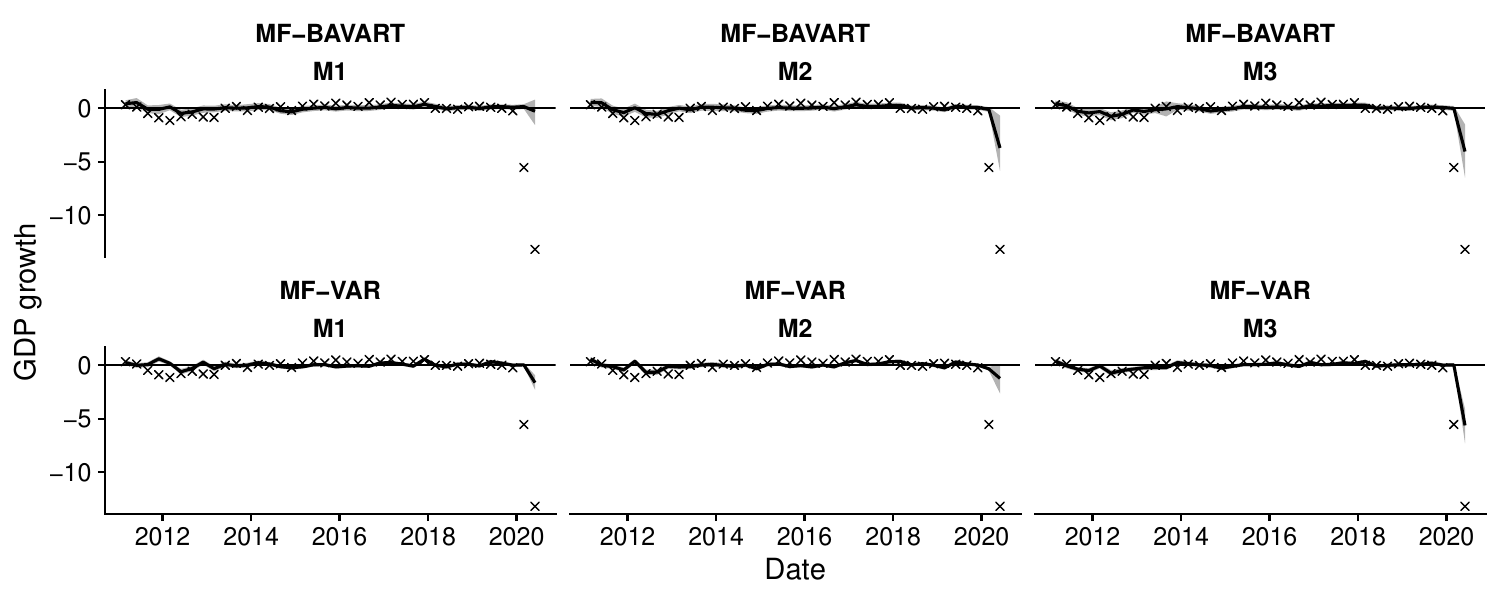}
\end{subfigure}
\caption*{\footnotesize\textit{Notes}: Columns are months per quarter in which the nowcast was produced. Realizations are marked as X's, and shown alongside the estimate for the posterior median and the 68 percent credible set.}
\end{figure}

\subsection{Estimates of the Latent States using FFBS compared to using BART}
In this sub-section, we briefly compare the posterior density obtained from using FFBS to the ones we obtain by using the estimated BART model. This serves to illustrate that both approaches (i.e. the one based on linearizing the non-linear model and the actual non-linear model) yield predictive distributions which are very similar. Notice that the main difference is that the FFBS-based posterior distribution features slightly tighter credible sets. These differences, however, only have a minor impact on the quality of the predictive density.

\begin{figure}[!ht]
\caption{Comparison of FFBS-based and BART-based estimates of the monthly GDP indicator.}\label{fig: pred_exact}
\centering
\begin{subfigure}[t]{.45\textwidth}
    \caption{DE}\label{pred_2020_GE_ffbs}
	\includegraphics[scale=0.35]{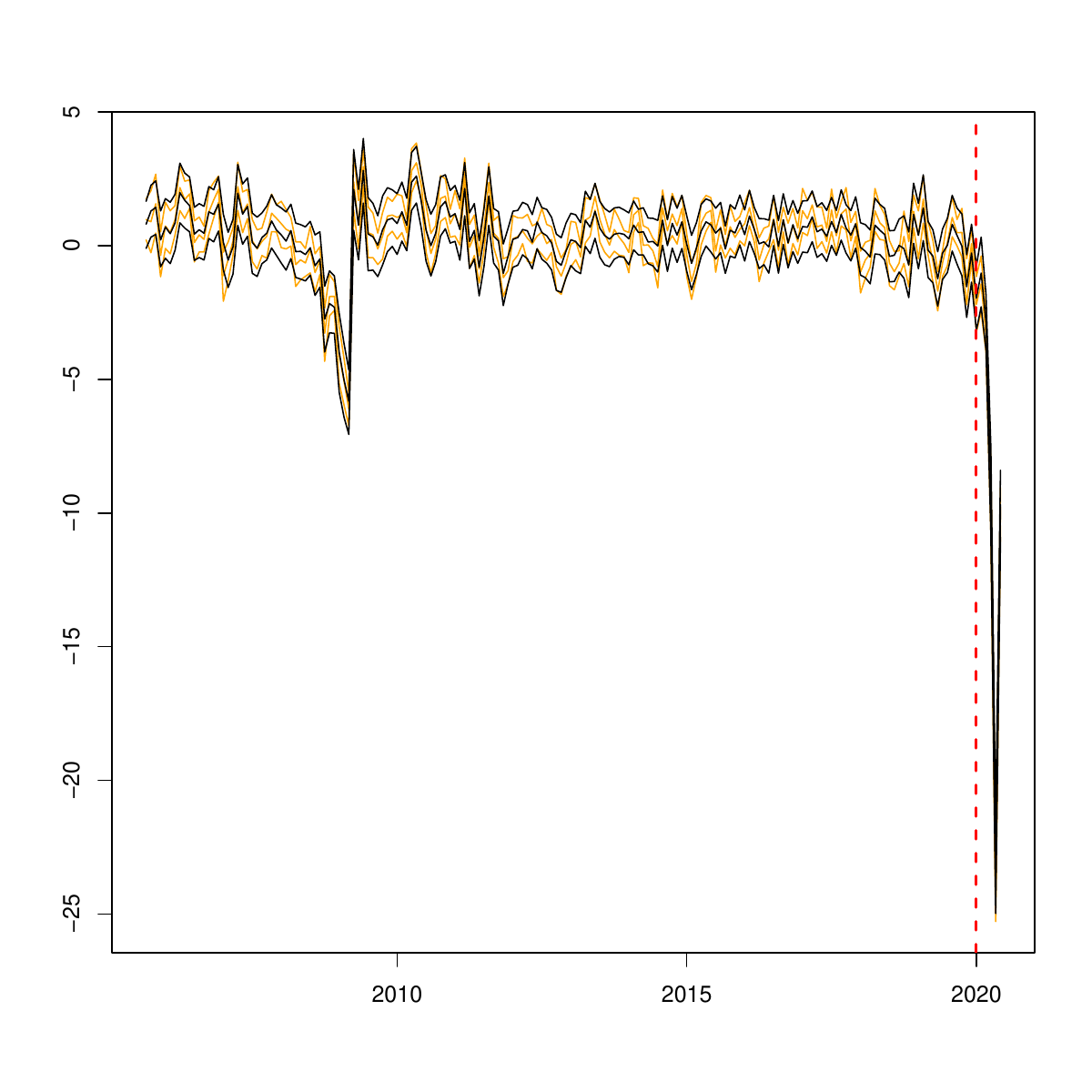}
\end{subfigure}
\begin{subfigure}[t]{.45\textwidth}
    \caption{ES}\label{pred_2020_ES_ffbs}
	\includegraphics[scale=0.35]{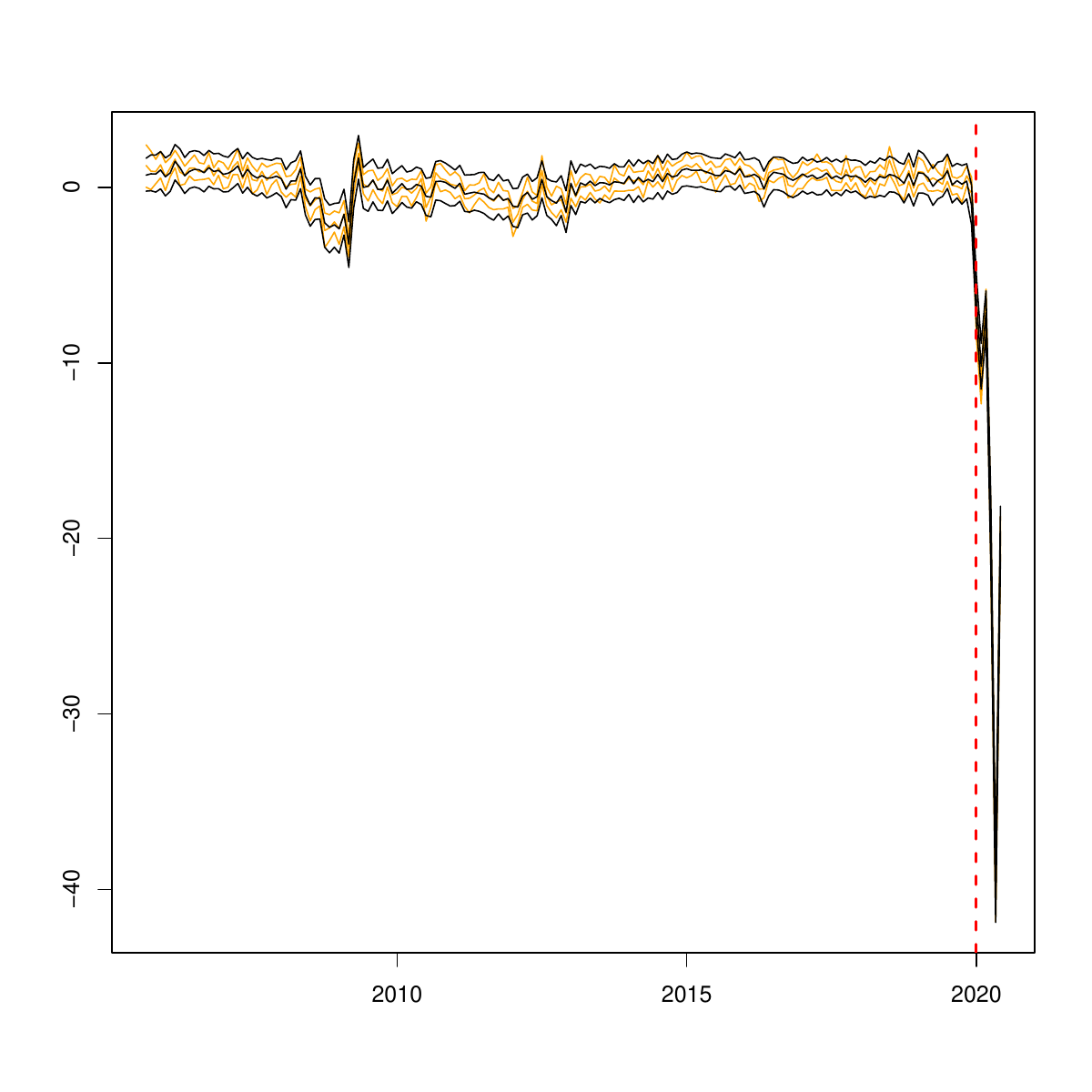}
\end{subfigure}
~
\begin{subfigure}[t]{.45\textwidth}
    \caption{FR}\label{pred_2020_FR_ffbs}
	\includegraphics[scale=0.35]{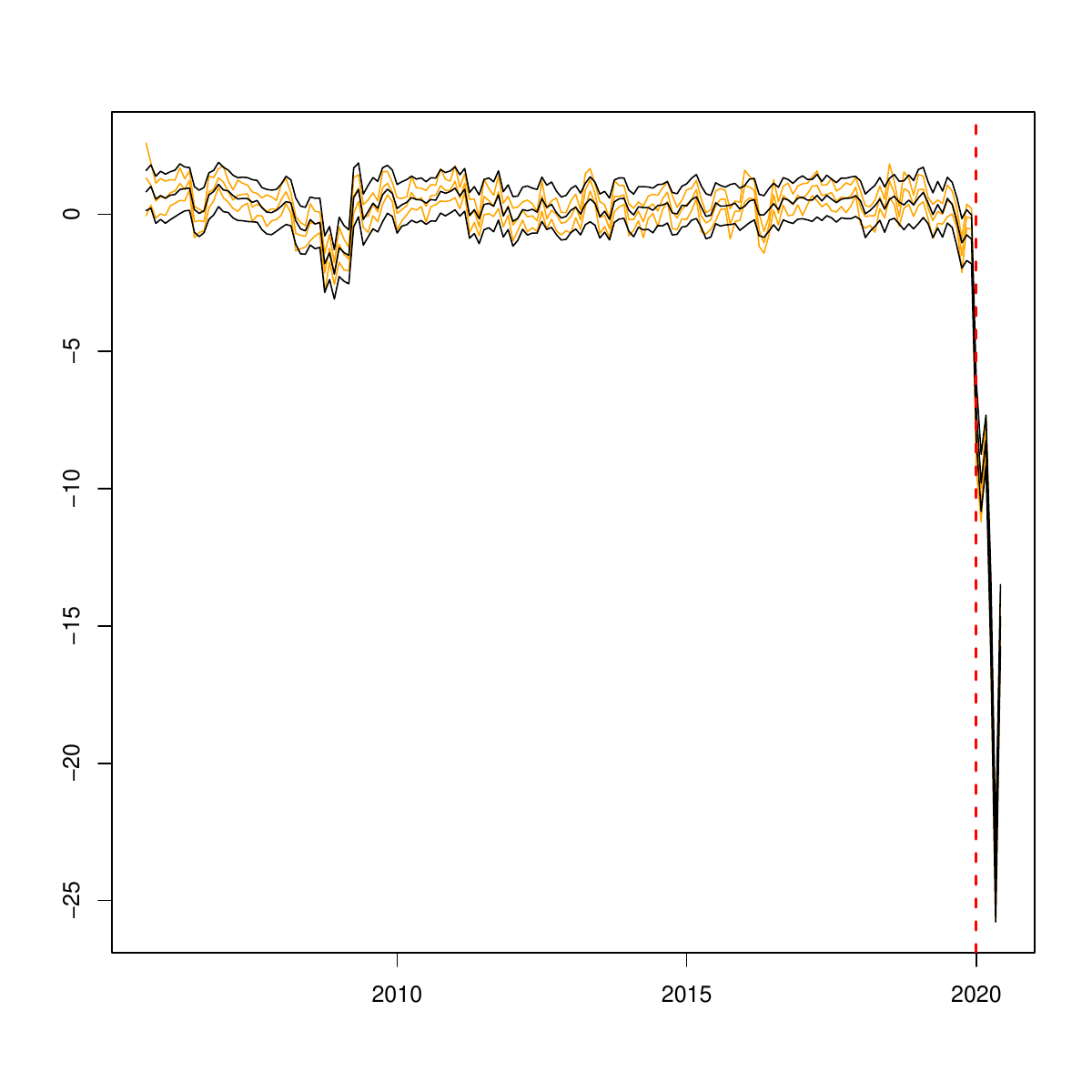}
\end{subfigure}
~
\begin{subfigure}[t]{.45\textwidth}
    \caption{IT}\label{pred_2020_IT_ffbs}
	\includegraphics[scale=0.35]{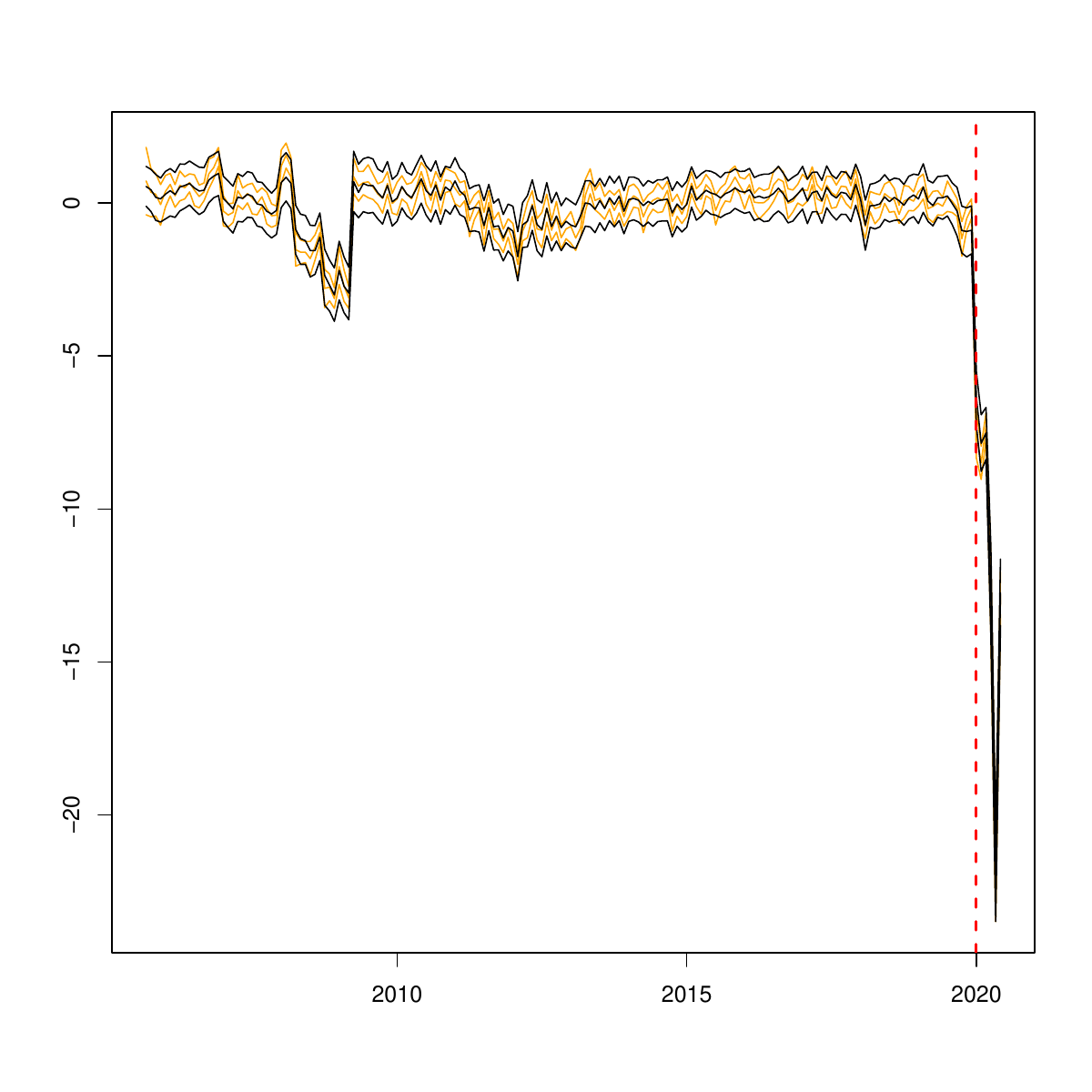}
\end{subfigure}
	\caption*{\footnotesize\textit{Notes}: The figure shows the 16$^\text{th}$, 50$^\text{th}$ and 84$^\text{th}$ precentiles of the posterior distribution of the latent monthly GDP indicators using Forward-Filtering Backward-Sampling (FFBS, in orange) and based on the estimated BART specification (in black). The dashed red line marks the beginning of the pandemic (March 2020).}
\end{figure}

\clearpage
\renewcommand\thesection{B}
\renewcommand\thetable{B.\arabic{table}}
\setcounter{table}{0}

\section{MCMC Diagnostics}
To assess convergence properties of our sampling algorithm, we present inefficiency factors (IF) and the \citet{rafterylewis1992} diagnostic (RL) of the total number of runs required to achieve a certain level of precision. The quantiles for the latter are set to $0.025$, the degree of accuracy is $0.025$, and the probability of attaining this required accuracy is $0.95$.

Table \ref{MCMCdiag} shows the corresponding measures across all countries for the final data vintage. For $\tilde{\bm{A}}$ we compute the median across all coefficients, $\det(\bm{\Sigma})$ is the determinant of the covariance matrix. For $\bm{y}_{q,\text{2005:2019}}$, we present the median over time from 2005M03 to 2019M12, while for $\bm{y}_{q,\text{2020}}$ we show the median over time from 2020M01 to 2020M06.

IFs and RLs signal an overall strong performance of the algorithm. Inefficiency factors are below five for all parameters and latent states except for $\bm{y}_{q,\text{2020}}$ in Spain. Considering the RL diagnostic, we find that the number of iterations required is far below the total number of iterations considered for most parameters. Overall the convergence statistics indicate satisfactory performance of our algorithm.

\begin{table*}[ht]
\caption{Summary statistics of MCMC diagnostics.}\vspace*{-1.5em}
\begin{center}
\begin{threeparttable}
\begin{tabular*}{\textwidth}{@{\extracolsep{\fill}} lrrrrrrrr}
  \toprule
&  \multicolumn{2}{c}{\textbf{DE}} & \multicolumn{2}{c}{\textbf{ES}} & \multicolumn{2}{c}{\textbf{FR}} & \multicolumn{2}{c}{\textbf{IT}} \\ 
 \cmidrule(lr){2-3}\cmidrule(lr){4-5}\cmidrule(lr){6-7}\cmidrule(lr){8-9}
 & IF & RL & IF & RL & IF & RL & IF & RL \\ 
  \midrule
$\tilde{\bm{A}}$ & 2.224 & 3951 & 2.672 & 3940 & 2.693 & 3994 & 2.507 & 3983 \\ 
  $\det(\bm{\Sigma})$ & 4.058 & 4338 & 4.870 & 4340 & 4.708 & 4314 & 4.453 & 4458 \\ 
  $\bm{y}_{q,\text{2005:2019}}$ & 1.538 & 3802 & 1.353 & 3782 & 1.411 & 3761 & 1.554 & 3802 \\ 
  $\bm{y}_{q,\text{2020}}$ & 3.458 & 3792 & 5.308 & 8937 & 4.247 & 8258 & 4.545 & 11830 \\ 
   \bottomrule
\end{tabular*}
\begin{tablenotes}[para,flushleft]
\footnotesize{\textit{Notes}: For $\tilde{\bm{A}}$ we compute the median measures across all coefficients, for $\bm{y}_{q,\text{2005:2019}}$, we present the median over time from 2005M03 to 2019M12, while for $\bm{y}_{q,\text{2020}}$ we show the median over time from 2020M01 to 2020M06. IF and RL denote inefficiency factors and the \citet{rafterylewis1992} diagnostic, respectively.}
\end{tablenotes}
\end{threeparttable}
\end{center}
\label{MCMCdiag}
\end{table*}
\end{appendices}
\end{document}